\documentclass{aa}  

\usepackage{graphicx}
\usepackage{txfonts}
\usepackage{color}
\usepackage{xcolor}

\usepackage{natbib}
\bibpunct{(}{)}{;}{a}{}{,}

\newcommand{\ergcm}[1]{$\times 10^{#1}$ erg cm$^{-2}$ s$^{-1}$}

\newcommand{\ergs}[1]{$\times 10^{#1}$ erg s$^{-1}$}

\newcommand{\ohcm}[1]{$10^{#1}$ cm$^{-2}$}

\newcommand{\kms}{km s$^{-1}$\xspace}

\newcommand{\cts}{cts s$^{-1}$\xspace}
\newcommand{\ftocr}[1]{$\times 10^{#1}$ erg cm$^{-2}$ cts$^{-1}$\xspace}

\newcommand{\Hone}{\ion{H}{i}\xspace}

\newcommand{\Halpha}{H${\alpha}$\xspace}
\newcommand{\Hbeta}{H${\beta}$\xspace}
\newcommand{\ltsima}{$\buildrel < \over \sim$}
\newcommand{\lsim}{\lower.5ex\hbox{\ltsima}}
\newcommand{\gtsima}{$\buildrel > \over \sim$}
\newcommand{\gsim}{\lower.5ex\hbox{\gtsima}}

\newcommand{\swift}{{\it Swift}\xspace}
\newcommand{\xmm}{{\it XMM-Newton}\xspace}

\newcommand{\rosat}{{\it ROSAT}\xspace}
\newcommand{\cxo}{{\it Chandra}\xspace}

\newcommand{\xspec}{{\tt XSPEC}\xspace}
\newcommand{\cutoffpl}{{\tt cutoffpl}\xspace}
\newcommand{\highecut}{{\tt highecut}\xspace}
\newcommand{\fdcut}{{\tt fdcut}\xspace}
\newcommand{\arfgen}{{\tt arfgen}\xspace}
\newcommand{\rmfgen}{{\tt rmfgen}\xspace}
\newcommand{\evselect}{{\tt evselect}\xspace}
\newcommand{\epiclccorr}{{\tt epiclccorr}\xspace}

\newcommand{\xmma}{\mbox{XMMU\,J045315.1$-$693242}\xspace}
\newcommand{\xmmb}{\mbox{XMMU\,J045736.9$-$692727}\xspace}
\newcommand{\rxjc}{\mbox{RX\,J0524.2$-$6620}\xspace}
\newcommand{\sxmma}{\mbox{J0453}\xspace}
\newcommand{\sxmmb}{\mbox{J0457}\xspace}
\newcommand{\srxjc}{\mbox{J0524}\xspace}
\newcommand{\massa}{\mbox{04531503$-$6932416}\xspace}
\newcommand{\massb}{\mbox{04573695$-$6927275}\xspace}
\newcommand{\massc}{\mbox{05241180$-$6620512}\xspace}

\begin{document} 

\title{Three new high-mass X-ray binaries in the Large Magellanic Cloud}

\author{F. Haberl\inst{\ref{mpe}} \and
        C. Maitra\inst{\ref{mpe}} \and
        G. Vasilopoulos\inst{\ref{oas}} \and
        P. Maggi\inst{\ref{oas}} \and
        A. Udalski\inst{\ref{uw}} \and
        I. M. Monageng \inst{\ref{saao}, \ref{uct}} \and
        D. A. H. Buckley\inst{\ref{saao}, \ref{uct}}
       } 

\titlerunning{Three new high-mass X-ray binaries in the LMC}
\authorrunning{Haberl et al.}

\institute{Max-Planck-Institut f{\"u}r extraterrestrische Physik, Gie{\ss}enbachstra{\ss}e 1, 85748 Garching, Germany\label{mpe}, \email{fwh@mpe.mpg.de}
\and
Universit\'e de Strasbourg, CNRS, Observatoire astronomique de Strasbourg, UMR 7550, F-67000 Strasbourg, France\label{oas}
\and
Astronomical Observatory, University of Warsaw, Al. Ujazdowskie 4, 00-478, 
Warszawa, Poland\label{uw}
\and
South African Astronomical Observatory, P.O. Box 9, Observatory, Cape Town 7935, South Africa\label{saao}
\and
Department of Astronomy, University of Cape Town, Private Bag X3, 7701 Rondebosch, South Africa\label{uct}
}

\date{Received 10 February 2022 / Accepted 1 March 2022}

\abstract
   {The Magellanic Clouds host a large population of high-mass X-ray binary (HMXB) systems, but although the Large Magellanic Cloud (LMC) is an order of magnitude more massive than the Small Magellanic Cloud, significantly fewer HMXBs are known.} 
   {We conducted a search for new HMXBs in \xmm observations, which we performed to investigate supernova remnant candidates in the supergiant shells LMC5 and LMC7.
   The three observed fields are located in regions, which were little explored in X-rays before.}
   {We analysed the \xmm data to look for sources with hard X-ray spectrum and counterparts with optical colours and brightness typical for HMXBs.}
   {We report the discovery of three new Be/X-ray binaries, two of them showing pulsations in their X-ray flux. With a luminosity of 6.5\ergs{34}, \xmma in LMC7 was relatively X-ray faint. The long-term OGLE I-band light curve of the V = 15.5\,mag counterpart suggests a 49.6\,day or 24.8\,day orbital period for the binary system. \xmmb, also located in LMC7 was brighter with a luminosity of 5.6\ergs{35} and hard spectrum with a power-law photon index of 0.63. The X-ray flux revealed clear pulsations with a period of 317.7\,s. We obtained optical high resolution spectra from the  V = 14.2\,mag counterpart using the SALT-HRS spectrograph. 
   \Halpha and \Hbeta were observed in emission with complex line profiles and equivalent widths of \hbox{-8.0}\,\AA\ and \hbox{-1.3}\,\AA, respectively.
   The I-band light curve obtained from OGLE shows a series of four strong outbursts followed by a sudden drop in brightness by more than 1\,mag within 73--165\,days and a recovery to the level before the outbursts. \rxjc, previously classified as X-ray binary candidate, is located at the eastern part of LMC5. We report the discovery of 360.7\,s pulsations. During the \xmm observation the luminosity was at $\sim$4\ergs{35} and the source showed a hard spectrum with a power-law photon index of 0.78. 
   The \Halpha emission line profile obtained from SALT-HRS is characterised by two broad peaks with a separation corresponding to $\sim$178\,\kms, and an equivalent width of -4.2\,\AA. The long-term OGLE I-band light curve of the V = 14.9\,mag counterpart reveals a quasi-periodic flaring activity while the colour evolution during the flares follows a hysteresis loop with redder colour during the rise.
   From modelling the \Halpha line profiles measured from \xmmb and \rxjc we derive constraints on the size of the Be disks.}
   {Our discovery of two pulsars among three new Be/X-ray binaries increases the number of known HMXB pulsars in the LMC to 25.}

\keywords{galaxies: individual: LMC --
          X-rays: binaries --
          stars: emission-line, Be -- 
          stars: neutron
         }

\maketitle   


\section{Introduction}
\label{sec:intro}

The Magellanic Clouds are well known for their large number of high-mass X-ray binaries (HMXBs).
In particular, the Small Magellanic Cloud (SMC) hosts more than 120 of these systems, 
nearly all of type Be/X-ray binary \citep[BeXRBs; see][for a review]{2011Ap&SS.332....1R}, 
in which a compact object orbits a Be star.
In most cases the compact object is a magnetised neutron star (NS), which accretes matter from the circum-stellar disk of the Be star and powers the X-ray emission.
For about half of the BeXRBs in the SMC pulsations in the X-ray flux were detected, 
which indicate the spin period of the NS \citep{2016A&A...586A..81H}.
The number of HMXBs known in the Large Magellanic Cloud (LMC) is smaller than in the SMC.
With the recent discoveries of a 570\,s BeXRB pulsar likely associated with a  supernova remnant \citep{2021MNRAS.504..326M} and 40.6\,s from the new BeXRB  eRASSU\,J050810.4$-$660653 \citep{2021ATel15133....1H}, 
the number of published LMC HMXB pulsars increased to 23, while additional $\sim$35 candidate HMXBs are known.
The star formation histories of LMC \citep{2016MNRAS.459..528A} and SMC \citep{2010ApJ...716L.140A} play an important role for the different number of active HMXBs, which we see today. The BeXRBs in the SMC are predominantly observed in regions, which experienced star formation bursts about 25--60 Myr ago, a time scale consistent with the evolutionary age of the Be phenomenon \citep{2010ApJ...716L.140A}.
Another reason for the lower number of HMXBs in the LMC is an observational bias due to the large extent of the galaxy on the sky which allowed deep mapping with sensitive X-ray instruments only of more central regions \citep[][]{2016A&A...585A.162M}.

We performed \xmm programmes (PI Maggi) to observe the hot interstellar medium, supernova remnant candidates, and HMXB candidates in the supergiant shells (SGSs) LMC5 (near the rim to LMC4) 
and LMC7 \citep{1980MNRAS.192..365M}.
These fields with young stellar populations are little explored and new HMXBs are expected to be found. To look for new HMXBs, we investigated the detected X-ray point sources with hard X-ray spectrum.
Many HMXB pulsars in the LMC are known in or near the rim of LMC4 \citep[see e.g.][]{2003A&A...406..471H}. The SGS LMC4 hosts the supergiant system LMC\,X-4 and six BeXRB pulsars.
Two BeXRBs in LMC7 were found in the past, the 187\,s pulsar Swift\,J045106.8$-$694803 by \citet{2013MNRAS.428.3607K} and 4XMM\,J045546.0$-$695717 \citep[Be cand 1 in][]{2018MNRAS.475.3253V}.
Here we present the discovery of a BeXRB pulsar with 360.7\,s period in LMC5, that is likely associated with \rxjc, proposed as X-ray binary by \citet{{2002A&A...388..100K}}.
We also report two new BeXRBs located in the SGS LMC7, one of them is a 317.7\,s pulsar. 
In Sect.\,\ref{sec:observations} we describe the analysis of the \xmm observations, the high-resolution optical spectra obtained at the Southern African Large Telescope \citep[SALT, ][]{Buckley2006}, and the archival monitoring data from the Optical Gravitational Lensing Experiment \citep[OGLE;][]{2008AcA....58...69U,2015AcA....65....1U}. We discuss the results in Sect.\,\ref{sec:discussion}.

\section{Observational data}
\label{sec:observations}

The two new HMXBs, which are located in the supergiant shell LMC7, were covered by two \xmm observations from our project with proposal ID 080455 (PI Maggi).
An additional, more recent serendipitous observation (0841660301) available in the \xmm archive was highly affected by background flaring activity. Because of the $\sim$50\% lower EPIC-pn exposure and the lower source flux (see below), we do not use this observation to determine the source position.
Our third new HMXB was found in the data of observation 0841320201 (MCSNR\,J0524-6624, PI Maggi).
Table\,\ref{tab:xmmobs} summarises the observations.

\subsection{X-ray positions and optical counterparts}
\label{sec:positions}

Following \citet{2013A&A...558A...3S}, we processed the data from the European Photon Imaging Camera (EPIC), which is equipped with pn- \citep{2001A&A...365L..18S} and 
MOS-type \citep{2001A&A...365L..27T} charge-coupled device (CCD) detectors.
We used the \xmm Science Analysis Software (SAS)\,19.1.0\footnote{\url{https://www.cosmos.esa.int/web/xmm-newton/sas}}
package and performed source detection simultaneously on the 15 images from the three EPIC instruments in five energy bands (0.2--0.5\,keV, 0.5--1\,keV, 1--2\,keV, 2--4.5\,keV, 4.5--12\,keV). 
This resulted in X-ray source positions of R.A. (2000) = 73.31102\degr, Dec. = -69.54548\degr\ for \xmma (\sxmma for short), R.A. (2000) = 74.40351\degr, Dec. = -69.45837\degr\ for \xmmb (\sxmmb) and R.A. (2000) = 81.04925\degr, Dec. = -66.34743\degr\ for \rxjc (\srxjc). Using background quasars located in the field of view (FoV) of the two \xmm observations of LMC7, we derived astrometric corrections 
of 2.7\arcsec--2.9\arcsec. 
No significant correction was required for \srxjc.
The final X-ray coordinates are listed in Table\,\ref{tab:xmmobs} and are used for the source names.
The position of our third source is 6.3\arcsec\ from a hard \rosat source (hardness ratio HR1=1.00$\pm$0.09) at
R.A. = 05$^{\rm h}$\,24$^{\rm m}$\,12.8$^{\rm s}$ and Dec. = $-$66\degr\,20\arcmin\,53\arcsec\ (J2000) with a 90\% confidence error of 12.6\arcsec, 
classified by \citet{2002A&A...388..100K} as X-ray binary.
Hence, we identify the \xmm source with \srxjc and use the \rosat name.

\begin{table*}
\centering
\caption[]{\xmm observations of \xmma, \xmmb and \rxjc.}
\label{tab:xmmobs}
\begin{tabular}{lclccrrr}
\hline\hline\noalign{\smallskip}
\multicolumn{1}{c}{Source} &
\multicolumn{1}{c}{Obs.} &
\multicolumn{1}{c}{Observation} &
\multicolumn{1}{c}{Exposure time} &
\multicolumn{1}{c}{Off-axis} &
\multicolumn{1}{c}{R.A.} &
\multicolumn{1}{c}{Dec.} &
\multicolumn{1}{c}{Err} \\
\multicolumn{1}{c}{name} &
\multicolumn{1}{c}{ID} &
\multicolumn{1}{c}{time} &
\multicolumn{1}{c}{pn, MOS1, MOS2} &
\multicolumn{1}{c}{angle} &
\multicolumn{2}{c}{(J2000)} &
\multicolumn{1}{c}{1$\sigma$} \\
\multicolumn{1}{c}{} &
\multicolumn{1}{c}{} &
\multicolumn{1}{c}{} &
\multicolumn{1}{c}{(s)} &
\multicolumn{1}{c}{(\arcmin)} &
\multicolumn{1}{c}{(h m s)} &
\multicolumn{1}{c}{(\degr\ \arcmin\ \arcsec)} &
\multicolumn{1}{c}{(\arcsec)} \\
\noalign{\smallskip}\hline\noalign{\smallskip}
\sxmma & 0804550201 & 2018-01-24 12:34 -- 21:44       & 28514, 0\tablefootmark{(a)}, 31494 & 6.8 & 04 53 15.11 & -69 32 42.5 & 0.7 \\
\sxmmb & 0804550101 & 2017-11-06 03:24 -- 13:17       & 32609, 34212, 34193                & 7.2 & 04 57 36.91 & -69 27 27.2 & 0.5 \\
\sxmmb & 0841660301 & 2020-04-11 21:38 -- (+1) 13:15  & 17200,  37635,               40068 & 5.2 &  --         & --          &  -- \\
\srxjc & 0841320201 & 2019-09-30 13:40 -- 21:40       & 22459,  26218,  26099              & 3.8 & 05 24 11.82 & -66 20 50.7 & 0.5 \\
\noalign{\smallskip}\hline
\end{tabular}
\tablefoot{
The net exposure times after background-flare screening are listed for pn, MOS1 and MOS2, respectively. 
The off-axis angle is given for the EPIC-pn telescope.
Observation 0841660301 was strongly affected by background flares, which significantly reduced the exposure time, in particular for EPIC-pn.
No flare was present during observation 0804550101, very few during 0804550201. 
These two observations were used to determine the source positions. 
A bore-sight correction using 12 and 8 QSOs in the FoV of observations 0804550101 and 0804550201 was applied to the source coordinates (see Sect.\,\ref{sec:positions}). The positional errors include the statistical and remaining systematic errors \citep[see][]{2013A&A...558A...3S}.
\tablefoottext{a}{\sxmma was located on a non-functional MOS1-CCD.}
}
\end{table*}

We found likely optical counterparts in the catalogues of the 
Magellanic Clouds Photometric Survey \citep{2004AJ....128.1606Z}
and the Two Micron All Sky Survey \citep[2MASS,][]{2003yCat.2246....0C}.
The magnitudes and colours are consistent with early type stars, suggesting a BeXRB nature for all three sources.
Table\,\ref{tab:optical} details some properties of the optical counterparts.

\begin{table*}
\centering
\caption[]{Optical counterparts of \xmma, \xmmb and \rxjc.}
\label{tab:optical}
\begingroup
\setlength{\tabcolsep}{4pt} 
\begin{tabular}{cccccccccc}
\hline\hline\noalign{\smallskip}
\multicolumn{1}{c}{Source} &
\multicolumn{1}{c}{V\tablefootmark{a}} &
\multicolumn{1}{c}{Q\tablefootmark{a,b}} &
\multicolumn{1}{c}{2MASS} &
\multicolumn{1}{c}{J} &
\multicolumn{1}{c}{H} &
\multicolumn{1}{c}{K$_{\rm s}$} &
\multicolumn{1}{c}{R.A.} &
\multicolumn{1}{c}{Dec.} &
\multicolumn{1}{c}{D\tablefootmark{d}} \\
\multicolumn{1}{c}{name} &
\multicolumn{6}{c}{} &
\multicolumn{2}{c}{(J2000)\tablefootmark{c}} &
\multicolumn{1}{c}{} \\
\multicolumn{1}{c}{} &
\multicolumn{1}{c}{(mag)} &
\multicolumn{1}{c}{(mag)} &
\multicolumn{1}{c}{} &
\multicolumn{1}{c}{(mag)} &
\multicolumn{1}{c}{(mag)} &
\multicolumn{1}{c}{(mag)} &
\multicolumn{1}{c}{(h m s)} &
\multicolumn{1}{c}{(\degr\ \arcmin\ \arcsec)} &
\multicolumn{1}{c}{(\arcsec)} \\
\noalign{\smallskip}\hline\noalign{\smallskip}
\sxmma & 15.49 & -0.85 & \massa & 15.51 & 15.19 & 15.27 & 04 53 15.05 & -69 32 41.5 & 0.97 \\
\noalign{\smallskip}
\sxmmb & 14.22 & -1.33 & \massb & 13.50 & 13.38 & 13.18 & 04 57 36.95 & -69 27 27.5 & 0.39 \\
\noalign{\smallskip}
\srxjc & 14.87 & -0.94 & \massc & 14.70 & 14.69 & 14.59 & 05 24 11.81 & -66 20 51.1 & 0.37 \\
\noalign{\smallskip}\hline
\end{tabular}
\endgroup
\tablefoot{
\tablefoottext{a}{Magnitudes and colours from \citet{2004AJ....128.1606Z}}
\tablefoottext{b}{Reddening-free parameter Q = U-B - 0.72(B-V). For the distribution of the Q parameter of BeXRBs in the SMC see \citet{2016A&A...586A..81H}.}
\tablefoottext{c}{Position of the optical counterpart from Gaia EDR3  \citep[see][]{2021A&A...649A...1G,2016A&A...595A...1G}.}
\tablefoottext{d}{Angular distance between \xmm and Gaia position.}
}
\end{table*}

\subsection{X-ray spectral analysis}
\label{subsec:spectra}

We removed times of increased flaring activity when the background was above a threshold of 8 counts ks$^{-1}$ arcmin$^{-2}$ for EPIC-pn and 2.5 counts ks$^{-1}$ arcmin$^{-2}$ for EPIC-MOS (7.0--15.0\,keV bands) and we extracted the EPIC X-ray spectra from circular regions around the source positions and nearby blank-sky areas for the background. The region sizes were adjusted to source brightness and off-axis angle.
From the EPIC-pn data single- and double-pixel events ({\tt PATTERN} 0--4) were selected,
excluding known bad CCD pixels and columns ({\tt FLAG} 0). Similarly, the filters {\tt PATTERN} 0--12 and {\tt FLAG} 0 were used for EPIC-MOS.
The spectra were re-binned to have at least twenty counts per bin, and $\chi^2$ statistic was used. The SAS tasks \arfgen and  \rmfgen were used to generate 
the corresponding detector response files. The X-ray spectra were analysed with the spectral fitting package 
\xspec\,12.11.0k\footnote{Available at \url{https://heasarc.gsfc.nasa.gov/xanadu/xspec/}} \citep{1996ASPC..101...17A}.
Errors are specified for 90\% confidence, unless otherwise stated.
A simple absorbed power-law model yielded acceptable spectral fits for \sxmma and \sxmmb, while for \srxjc a more complex model was required.
For the absorption we used two column densities along the line of sight.
One accounts for the Galactic foreground with solar abundances according to \citet{2000ApJ...542..914W} and was fixed at the value obtained from \Hone measurements \citep{1990ARA&A..28..215D}\footnote{Extracted using NASA's HEASARC web interface \url{https://heasarc.gsfc.nasa.gov/cgi-bin/Tools/w3nh/w3nh.pl}}, 
the other (free in the fit), with metal abundances set to 0.49 \citep{2002A&A...396...53R}, reflects the absorption by the interstellar medium of the LMC and local to the source.

\subsubsection{\xmma}
\label{subsec:speca}

\begin{figure}
  \centering
  \resizebox{0.95\hsize}{!}{\includegraphics[clip=]{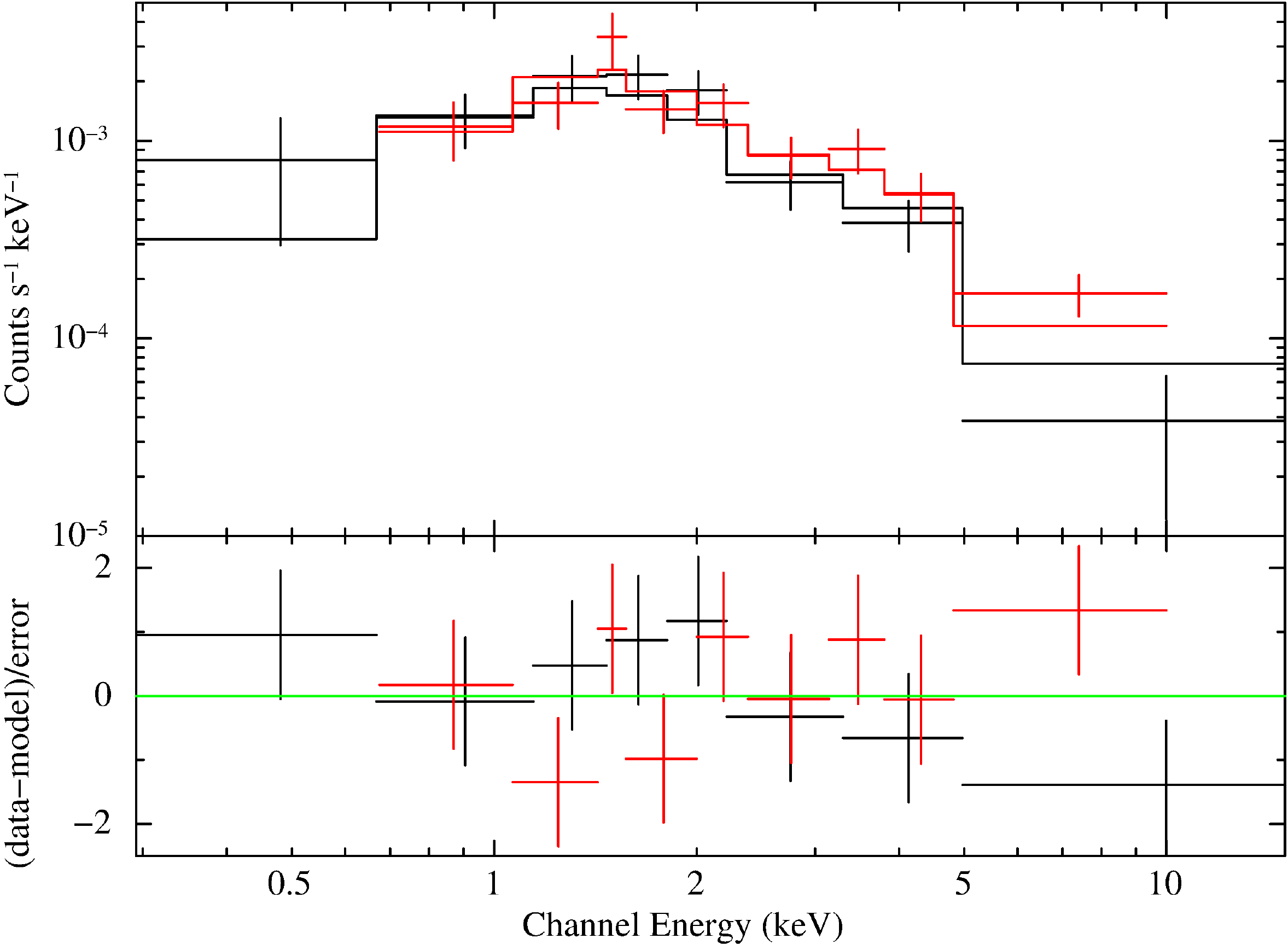}}
  \caption{
    EPIC spectra of \sxmma together with the best-fit absorbed power-law model (histograms).
    The EPIC-pn count rate (data and model shown in black) is reduced due to the source location on a chip gap. No MOS1 data are available, MOS2 data are marked in red.
    }
  \label{fig:EPICspeca}
\end{figure}

\sxmma was located on a chip gap of EPIC-pn, which significantly reduced the number of counts available for the analysis, moreover, no MOS1 data are available. 
We extracted counts within a circle around the source with radius 24\arcsec, while the background was taken from a nearby source-free region with radius of 42\arcsec. 
The EPIC-pn and -MOS2 spectra (118 and 167 net source counts, respectively) were fitted simultaneously with the same power-law model with all parameters linked, except a normalisation factor to allow for first-order cross calibration uncertainties between the EPIC instruments. The spectra are shown in Fig.\,\ref{fig:EPICspeca} together with the best-fit model. The model parameters, observed fluxes and source luminosities are summarised in Table\,\ref{tab:xspec}.

\begin{table*}
\centering
\caption[]{X-ray spectral fit results.}
\begin{tabular}{lccccccccc}
\hline\hline\noalign{\smallskip}
\multicolumn{1}{c}{Source} &
\multicolumn{1}{c}{Obs.} &
\multicolumn{1}{c}{Model} &
\multicolumn{1}{c}{Photon} &
\multicolumn{1}{c}{N$_{\rm H}^{\rm Gal}$} &
\multicolumn{1}{c}{N$_{\rm H}^{\rm LMC}$} &
\multicolumn{1}{c}{$\chi^2_{\rm r}$} &
\multicolumn{1}{c}{dof} &
\multicolumn{1}{c}{F$_{\rm observed}$\tablefootmark{a}} &
\multicolumn{1}{c}{L\tablefootmark{b}} \\
\multicolumn{1}{c}{name} &
\multicolumn{1}{c}{ID} &
\multicolumn{1}{c}{ } &
\multicolumn{1}{c}{index} &
\multicolumn{1}{c}{(\ohcm{21})} &
\multicolumn{1}{c}{(\ohcm{21})} &
\multicolumn{1}{c}{} &
\multicolumn{1}{c}{} &
\multicolumn{1}{c}{(erg cm$^{-2}$ s$^{-1}$)}&
\multicolumn{1}{c}{(erg s$^{-1}$)} \\
\noalign{\smallskip}\hline\noalign{\smallskip}
  \sxmma & 0804550201 & PL &  $1.21^{+0.44}_{-0.38}$ & 0.85  & $6.6^{+8.6}_{-5.9}$    & 1.0  & 13  & $1.85^{+0.17}_{-0.32} \times 10^{-13}$ &  $6.5 \times 10^{34}$ \\
  \noalign{\smallskip}\hline\noalign{\smallskip}
  \sxmmb & 0804550101 & PL &  $0.64 \pm 0.03$        & 0.85  & $<$0.17                & 0.98 & 388 & $1.83 \pm 0.05 \times 10^{-12}$        &  $5.6 \times 10^{35}$ \\
  \noalign{\smallskip}
  \sxmmb & 0841660301 & PL &  $0.63 \pm 0.05$        & 0.85  & $<$0.47                & 1.07 & 212 & $0.97 \pm 0.03 \times 10^{-12}$        &  $2.9 \times 10^{35}$ \\
  \noalign{\smallskip}
  \sxmmb & combined   & PL &  $0.64 \pm 0.03$        & 0.85  & $<$0.14                & 1.01 & 602 & -- &  -- \\
  \noalign{\smallskip}\hline\noalign{\smallskip}
  \srxjc\tablefootmark & 0841320201 & PL & $0.9^{+0.05}_{-0.05}$ &  0.54 & $2.5^{+0.5}_{-0.4}$    & 1.24  & 248  & $1.30^{+0.03}_{-0.04} \times 10^{-12}$ &  $3.9 \times 10^{35}$ \\
  \noalign{\smallskip}
  \srxjc\tablefootmark{c} & 0841320201 & \highecut & $0.78^{+0.06}_{-0.06}$ &  0.54 & $1.9^{+0.4}_{-0.5}$    & 1.06  & 246  & $1.24^{+0.03}_{-0.04} \times 10^{-12}$ &  $3.9 \times 10^{35}$ \\
  \noalign{\smallskip}
  \srxjc\tablefootmark{d} & 0841320201 & \fdcut & $0.78^{+0.06}_{-0.07}$ & 0.54   & $1.9^{+0.5}_{-0.4}$    & 1.07  & 246  & $1.25^{+0.03}_{-0.04} \times 10^{-12}$ &  $4.0 \times 10^{35}$ \\
  \noalign{\smallskip}
  \srxjc\tablefootmark{e} & 0841320201 & PL+BB & $2.3^{+0.5}_{-0.5}$ & 0.54   & $5.2^{+1.5}_{-1.3}$    & 1.02  & 246  & $1.26^{+0.03}_{-0.04} \times 10^{-12}$ &  $4.8 \times 10^{35}$ \\
\noalign{\smallskip}\hline

\end{tabular}
\tablefoot{
Best-fit parameters using a model with absorbed power-law emission. 
Errors indicate 90\% confidence ranges.
\tablefoottext{a}{Fluxes are provided for the 0.2--10.0\,keV band.}
\tablefoottext{b}{Source luminosities (0.2--10.0\,keV) corrected for absorption, assuming a distance of 50\,kpc \citep{2013Natur.495...76P}.} 
The Galactic foreground column density was taken from \citet{1990ARA&A..28..215D}. Fluxes and luminosities are taken as mean of the values from the different EPIC instruments.
\tablefoottext{c}{The PL is modified by an exponential cutoff (\highecut in \xspec) with cut-off energy at 6.5$^{+1.1}_{-0.5}$\,keV, and a folding energy of 4.1$^{+2.0}_{-0.8}$\,keV.}
\tablefoottext{d}{The PL is modified by a modified exponential cutoff (\fdcut, see text) with cut-off energy at 9.3$^{+0.6}_{-0.5}$\,keV, and a folding energy at 0.9$^{+0.5}_{-0.4}$\,keV.}
\tablefoottext{e}{The spectral model includes an additional black-body component with kT = 1.95$^{+0.15}_{-0.12}$\,keV and a size of 460$^{+60}_{-70}$\,m that accounts for $\sim$63\% of the flux in the 0.2-10.0 keV band.
}
}
\label{tab:xspec}
\end{table*}

\subsubsection{\xmmb}
\label{subsec:specb}

\begin{figure}
  \centering
  \resizebox{0.95\hsize}{!}{\includegraphics[clip=]{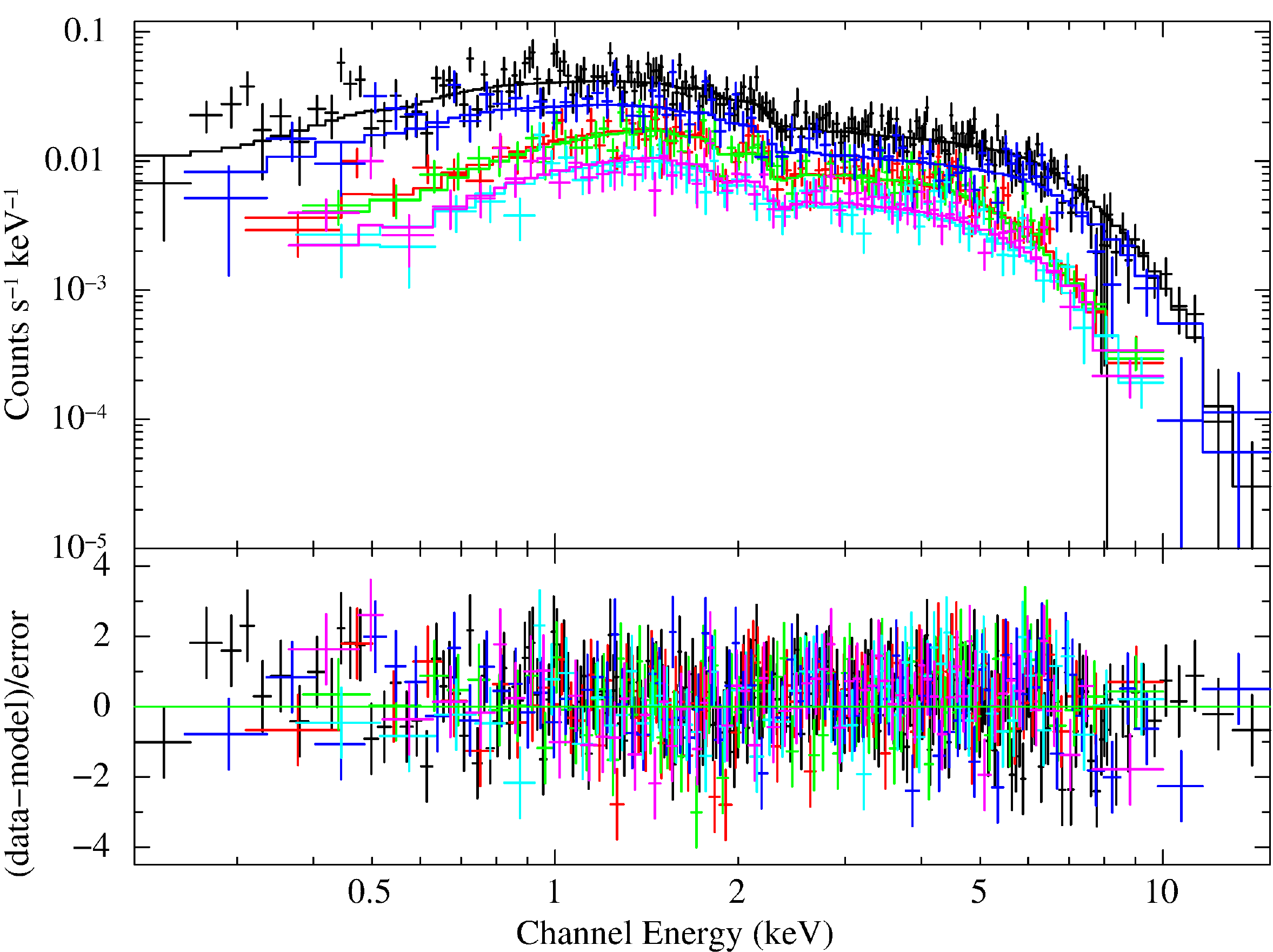}}
  \caption{
    EPIC spectra of \sxmmb from the two \xmm observations (Table\,\ref{tab:xmmobs}) together with the best-fit absorbed power law.
    The EPIC-pn, MOS1 and MOS2 data and best-fit model are shown in black (blue), green (light blue) and red (magenta) for obsID 0804550101 (0841660301), respectively.
    }
  \label{fig:EPICspecb}
\end{figure}

The position of \sxmmb was covered by all three EPIC instruments during both available \xmm observations, which allowed us to extract spectra with 4100 (1410), 1740 (1100) and 1685 (1270) source counts within a circle with radius 37\arcsec\ (30\arcsec) from the pn, MOS1 and MOS2 data of observation  0804550101 (0841660301), respectively. 
The background was taken from a nearby source-free region with radius of 44\arcsec\ for both observations.
We again fitted the spectra from the three instruments simultaneously with a power law  with individual normalisation factors, first keeping the two observations separately. The best-fit parameters (Table\,\ref{tab:xspec}) agree within their uncertainties. Hence, we combined in a second step all six spectra, which are presented in Fig.\,\ref{fig:EPICspecb}, together with the best-fit absorbed power-law model. Observed fluxes and source luminosities inferred from the individual observations are also listed in Table\,\ref{tab:xspec}.

\subsubsection{\rxjc}
\label{subsec:specc}

The position of \srxjc was also covered by all EPIC instruments during the \xmm observation (obsID 0841320201, see Table \ref{tab:xmmobs}). We extracted source (background) events from circular regions with radius 30\arcsec\ (50\arcsec). We fitted the spectra (2800, 1430 and 1550 net source counts from pn, MOS1 and MOS2, respectively) from the three instruments simultaneously with individual normalisation factors. In contrast to the previous two systems an absorbed single component model (i.e. power law) did not provide a good fit as was evident from the remaining features in the residuals. In particular the residuals in the high-energy part of the spectrum indicated some curvature. 
To improve the fit, we attempted to use modified power laws with cutoff or to add a second continuum component.
We first applied the standard \cutoffpl model in \xspec, which multiplies an exponential roll-off to the power law, while using one extra free parameter, the folding energy $E_f$ (i.e. $E^{-\Gamma}exp{(-E/E_f)}$). This only marginally improved the fit and the residuals at high energies remained. We thus opted for other modified power-law models, which have two additional free parameters. 
First, the \xspec model \highecut was multiplied to the power law above the cutoff energy $E_c$ ($E^{-\Gamma}exp{((E_c-E)/E_f)}$) and second, we used the 
\fdcut model of \citet{1986LNP...255..198T} that was successfully used to model spectra of accreting X-ray pulsars \citep[e.g.][]{2020MNRAS.494.5350V}. 
The latter model is a power law with smooth cutoff and is expressed analytically as:
\begin{equation}
dN/dE = E^{-\Gamma}\left(1+\exp{\frac{E-E_c}{E_f}}\right)^{-1}.   
\end{equation}
The absorbed \highecut and \fdcut models provide acceptable fits with similar $\chi^2$ (Table\,\ref{tab:xspec}).

In terms of models with two components,
a low temperature black-body component that could account for thermal emission from the accretion disk did not improve the fit nor fixed the residuals at higher energies.
Amongst two-component models, an acceptable fit was achieved by a combination of a power-law and a hot black-body component as it offered a better description of the data and reduced the residuals. The goodness of fit as measured by $\chi^2$ improved by 55.3 with the addition of two model parameters in comparison to the simple power-law model.
The temperature of the black-body component was found to be 1.95$^{+0.15}_{-0.12}$ keV while its size of the emission region of 460$^{+60}_{-70}$\,m appears to be consistent with a hot spot on the NS surface.  

\begin{figure}
  \centering
  \resizebox{0.95\hsize}{!}{\includegraphics[clip=]{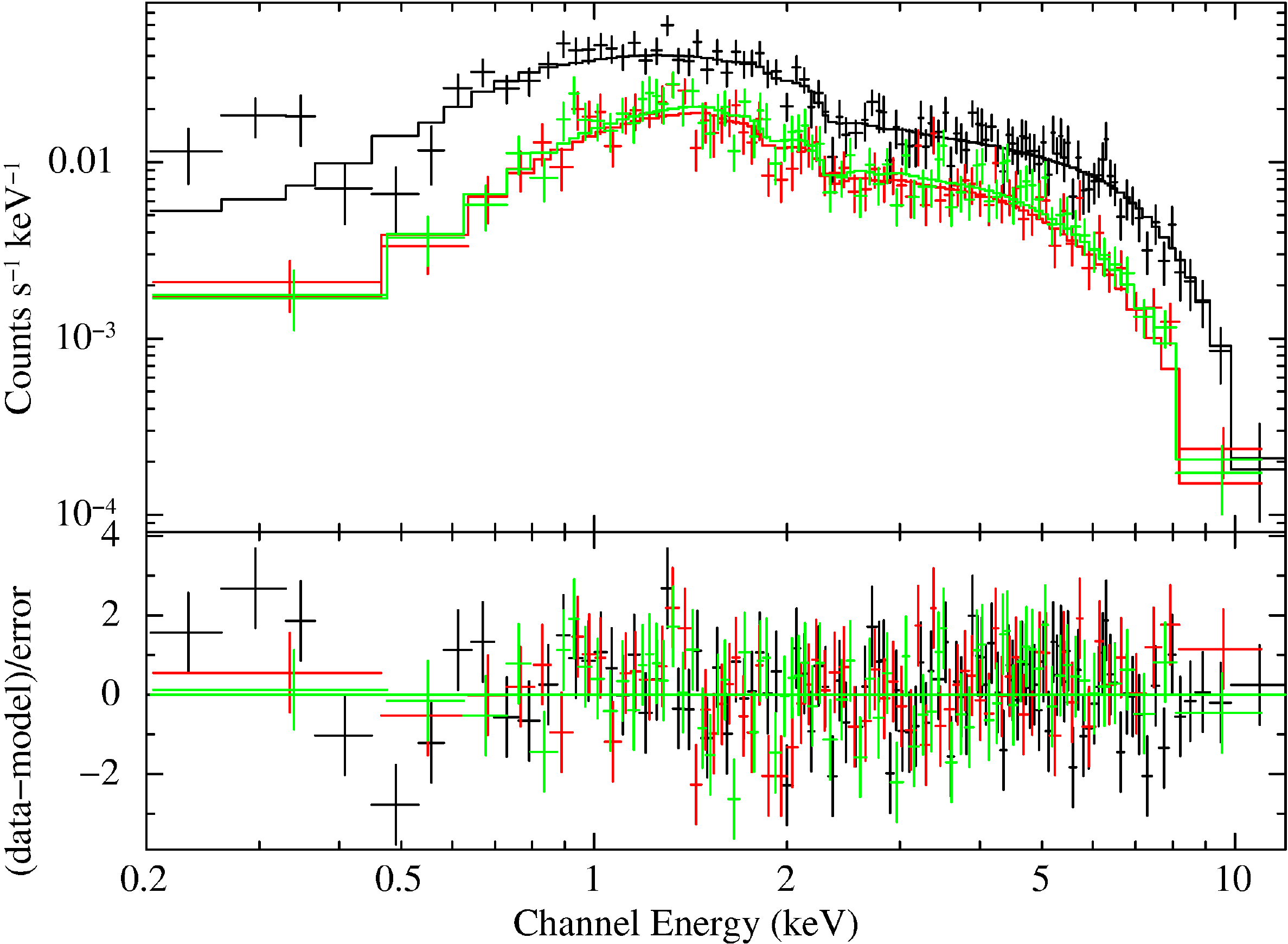}}
  \caption{
    EPIC spectra of \srxjc together with the best-fit absorbed power-law model with high-energy cutoff (\fdcut, see Table\,\ref{tab:xspec}) as histograms. 
    The EPIC-pn, MOS1 and MOS2 data and best-fit model are shown in black, green and red, respectively.
    }
  \label{fig:EPICspecc}
\end{figure}

\subsection{X-ray timing analysis}
\label{subsec:temp}

After correcting the event arrival times to the solar system barycentre, we created background subtracted X-ray light curves of \sxmma, \sxmmb and \srxjc in the 0.2--12\,keV band by using the SAS tasks \evselect and \epiclccorr with an initial binning of 5\,s. The same event selection criteria were used as for the spectra. 
We searched for a periodic signal in the \xmm EPIC light curves using a Lomb-Scargle (LS) periodogram analysis \citep{1976Ap&SS..39..447L,1982ApJ...263..835S} in the range of 10--3000\,s, which covers typical periods seen from HMXB pulsars.

\subsubsection{\xmma}
\label{subsec:tempa}

\begin{figure}
  \centering
  \resizebox{0.95\hsize}{!}{\includegraphics[]{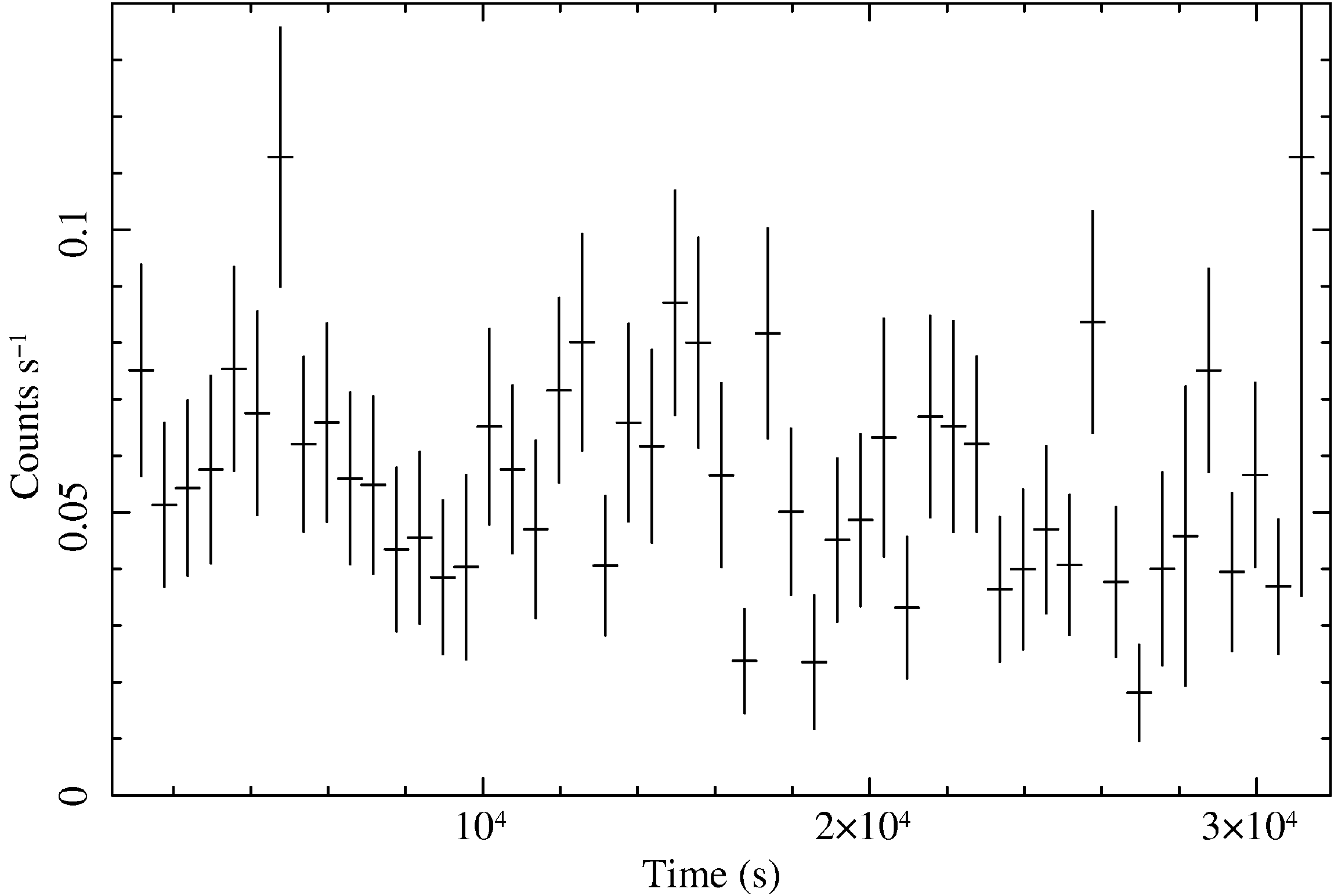}}
  \caption{
    Background-subtracted X-ray light curve of \sxmma in the 0.2--12\,keV energy band with time bins of 600\,s.
    The light curve is vignetting and PSF corrected and combines EPIC-pn and MOS2 data. Time zero corresponds to MJD 58142.5517 (2018 January 24, 13:19:28.7 UT).
  }
  \label{fig:epiclca}
\end{figure}

The X-ray light curve of \sxmma rebinned to 600\,s is presented in Fig.\,\ref{fig:epiclca}. Some variations in the flux are visible, but the large errors don't allow us to draw more quantitative conclusions. Fitting a constant to the light curve yields a reduced $\chi^2$ of 1.5 for 50 degrees of freedom.

\subsubsection{\xmmb}
\label{subsec:tempb}

\begin{figure}
  \centering
  \resizebox{0.95\hsize}{!}{\includegraphics[]{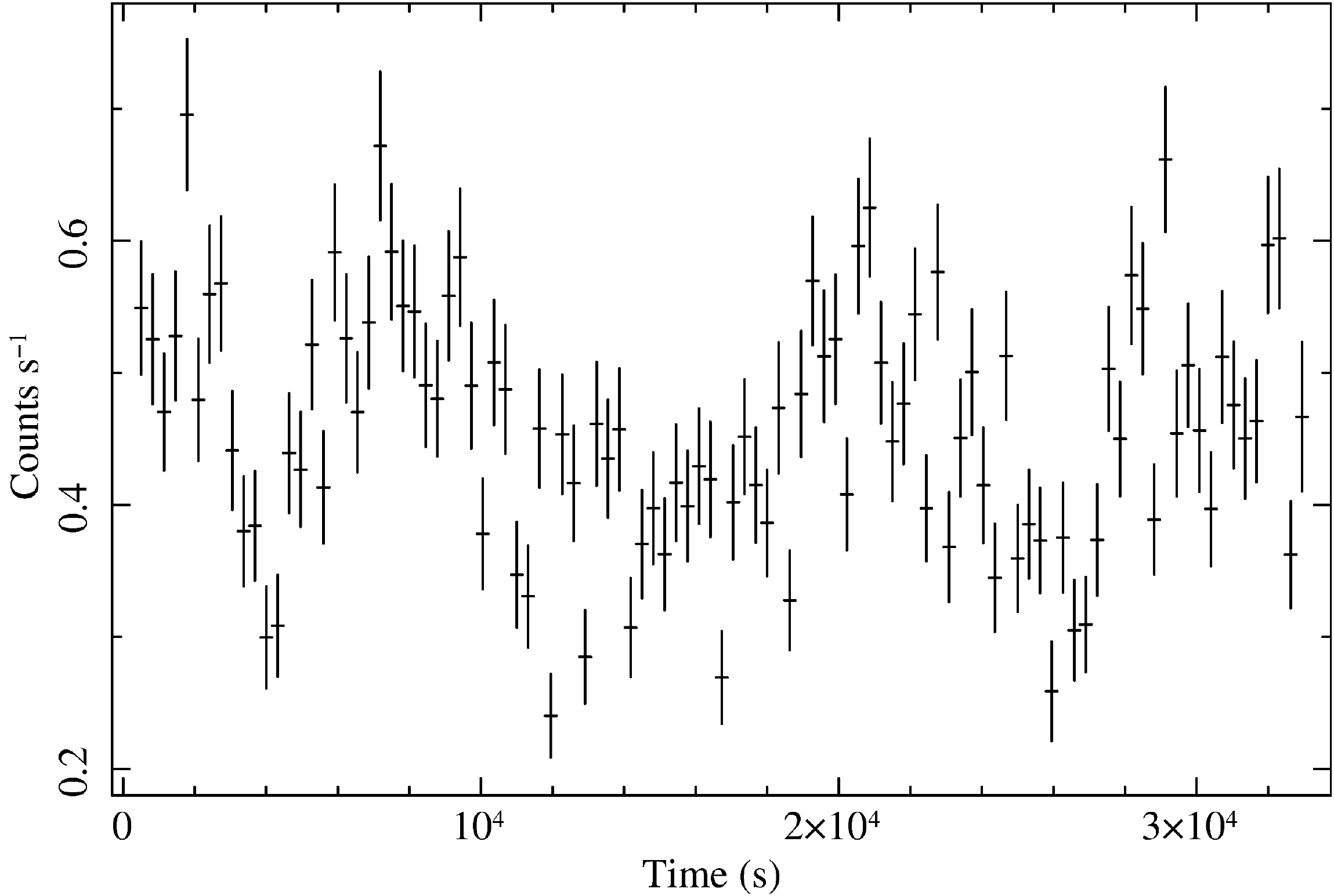}}
  \caption{
    Background-subtracted X-ray light curve of \sxmmb from observation 0804550101 in the 0.2--12\,keV energy band with time bins of 318\,s (corresponding to the pulse period).
    The light curve is vignetting and PSF corrected and combines EPIC-pn, MOS1 and MOS2 data. Time zero corresponds to MJD 58063.1707 (2017 November 6, 04:05:50.6 UT).
  }
  \label{fig:epiclcb}
\end{figure}

\begin{figure}
  \centering
  \resizebox{\hsize}{!}{\includegraphics[clip=]{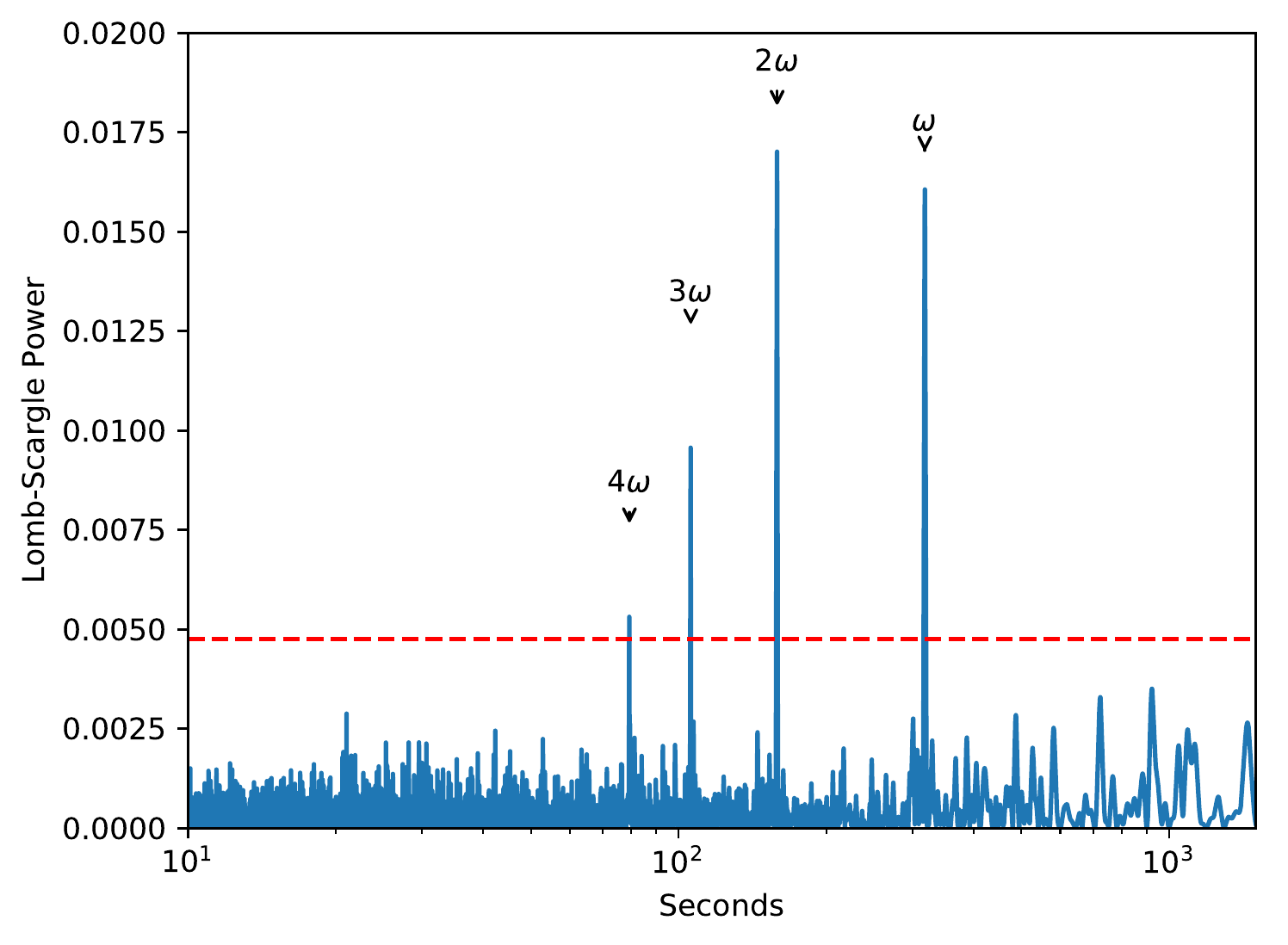}}
  \caption{
    LS periodogram of \sxmmb. Pulsations are clearly detected with a fundamental frequency corresponding to a period of $\sim$318\,s together with signals at three harmonic frequencies. The red dashed line marks the 3$\sigma$ confidence level.
  }
  \label{fig:XMMLSb}
\end{figure}

\begin{figure}
  \centering
  \resizebox{0.95\hsize}{!}{\includegraphics[]{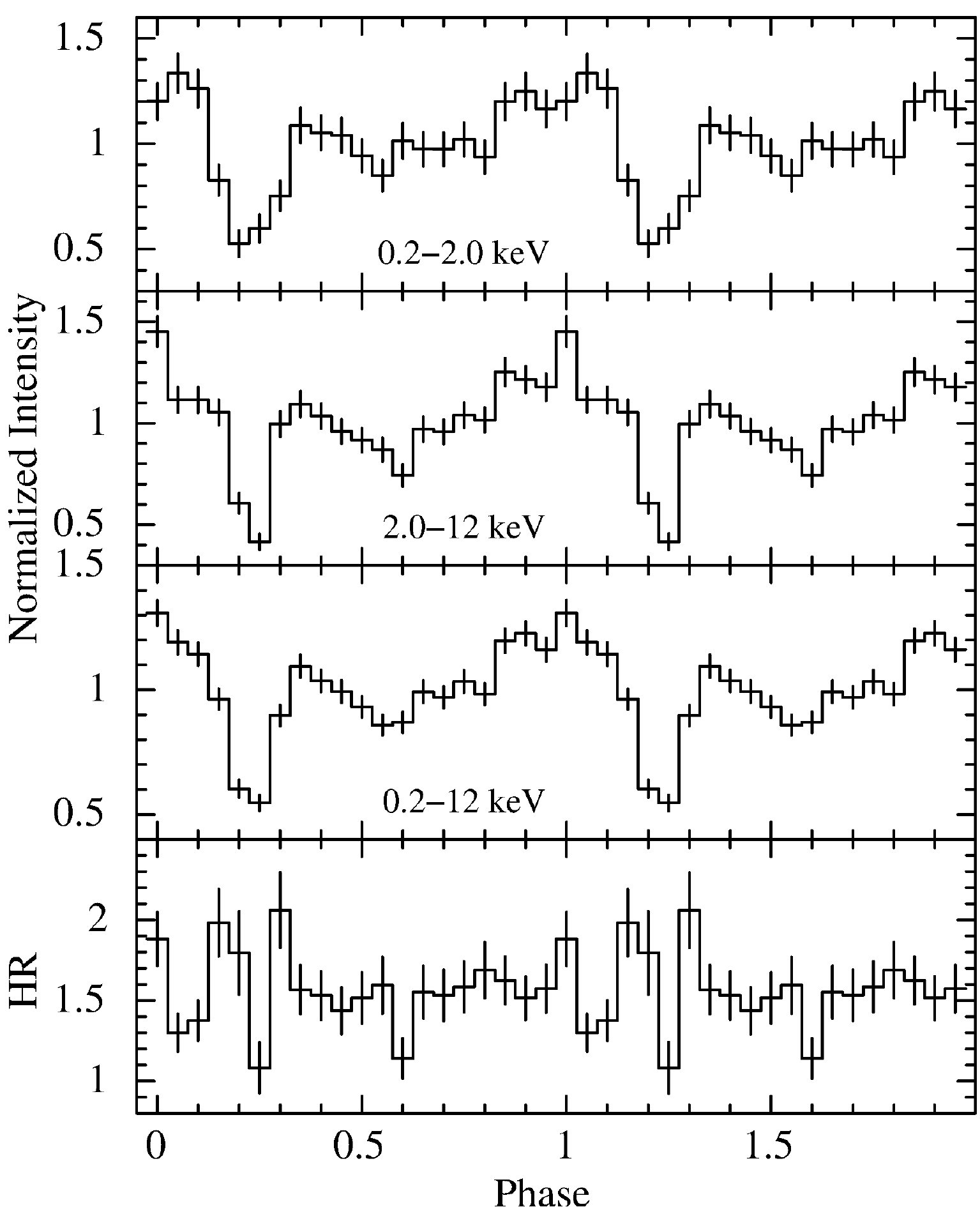}}
  \caption{
    Corrected EPIC-pn light curves folded with 317.72\,s showing the pulse profile of \sxmmb in the energy bands of 0.2--2.0\,keV (soft), 2.0--12\,keV (hard) and total 0.2--12\,keV. 
    The intensity profiles are normalised to their mean count rates, 0.189\,\cts, 0.268\,\cts and 0.457\,\cts for the soft, hard and total energy bands, respectively.
    For the hardness ratio (HR) the count rate in the hard band was divided by that in the soft band.
    A deep, sharp dip can be seen, which causes the high number of harmonic frequency peaks in the periodogram (Fig.\,\ref{fig:XMMLSb}).
  }
  \label{fig:ppb}
\end{figure}

The X-ray light curve of \sxmmb from the November 2017 observation, rebinned to 318\,s (the pulse period, see below) using the combined EPIC data shows flux variations by about a factor of two on time scales of typically one hour (Fig.\,\ref{fig:epiclcb}).
A strong periodic signal is detected with a fundamental frequency corresponding to $\sim$318\,s together with three harmonics as shown in Fig.~\ref{fig:XMMLSb}. 
The fundamental frequency indicates the spin frequency of the accreting NS, while the large number of harmonics suggests a complex pulse profile.
In order to determine the pulse period more precisely and its uncertainty, we employed the Bayesian periodic signal detection method described by \citet{1996ApJ...473.1059G}.
The inferred spin period with its associated 1$\sigma$ error are  317.72$\pm$0.05\,s.
The background subtracted \xmm EPIC light curves, folded with the best-obtained period, are characterised by a narrow deep dip as shown in Fig.~\ref{fig:ppb}.

Despite the strong background flaring activity during the April 2020 observation we searched for the pulse period.
We used data filtered and not filtered for the background flares. Both methods led to a very noisy LS periodogram, either due to the many data gaps or due to the highly variable background. We were able to determine the period with the Bayesian method to 317.08$\pm$0.14\,s, indicating a small spin-up at $\sim$4.3$\sigma$ confidence.

\begin{figure}
  \centering
  \resizebox{0.95\hsize}{!}{\includegraphics[]{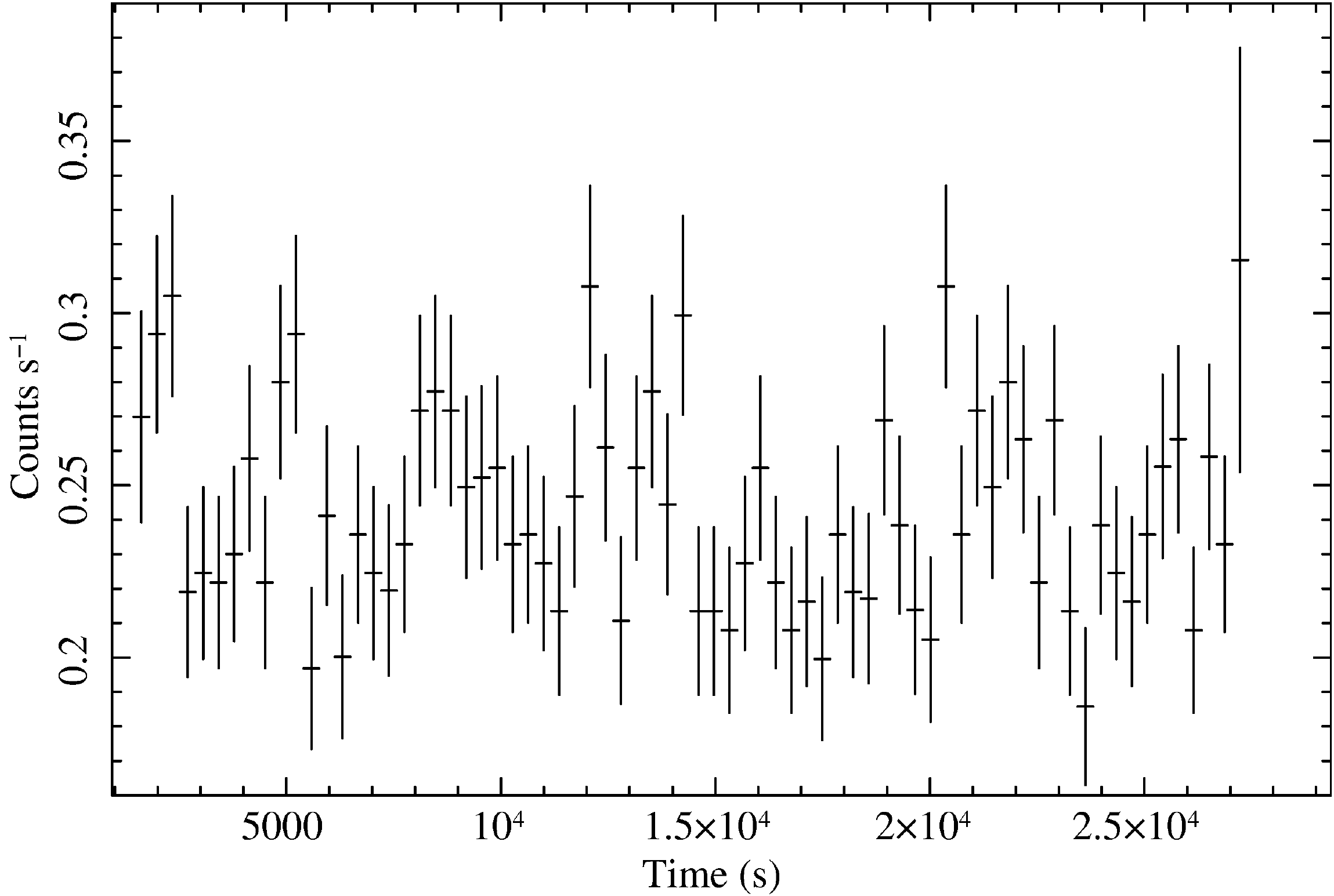}}
  \caption{
    Background-subtracted X-ray light curve of \srxjc in the 0.2--12\,keV energy band with time bins of 630.7\,s (corresponding to the pulse period).
    The light curve is vignetting and PSF corrected and combines EPIC-pn, MOS1 and MOS2 data. 
  }
  \label{fig:epiclcc}
\end{figure}


\begin{figure}
  \centering
  \resizebox{0.95\hsize}{!}{\includegraphics[]{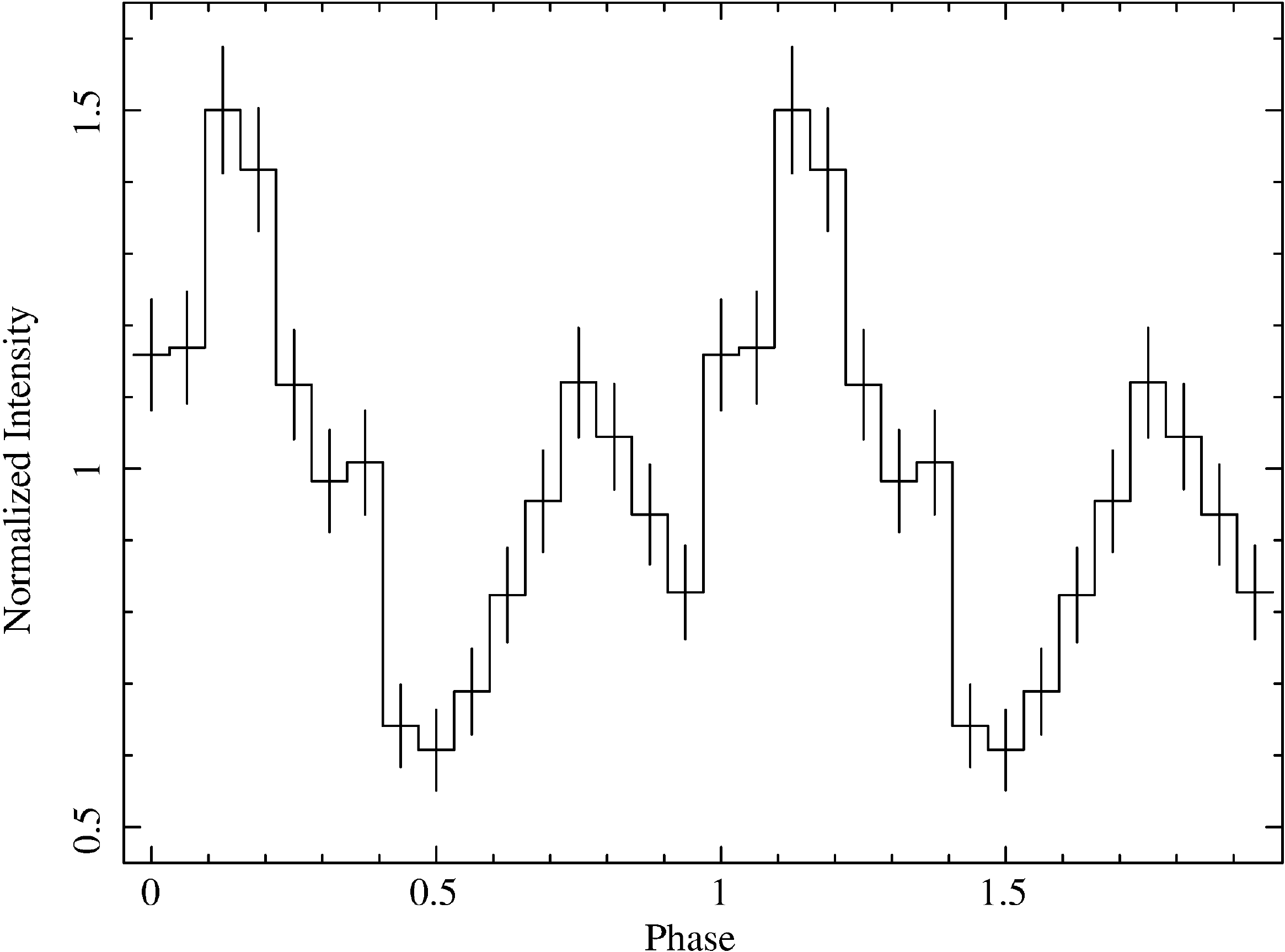}}
  \caption{
    Corrected EPIC-pn light curve folded with 360.7\,s showing the pulse profile of \srxjc in the energy band of 0.2--12\,keV,
    normalised to the mean count rate of  0.124\,\cts.
  }
  \label{fig:ppc}
\end{figure}

\subsubsection{\rxjc}
\label{subsec:tempc}

To investigate the temporal properties of the system we followed similar procedure as with the other two pulsars. A pulse period of 360.70$\pm0.12$\,s was determined following the Bayesian periodic signal detection method described by \citet{1996ApJ...473.1059G}.
The X-ray light curve derived from all EPIC cameras is plotted in Fig. \ref{fig:epiclcc}. No significant variability on time scales larger than the spin period is evident from the data.
The pulse profile is double peaked while the major and minor peaks are separated by 0.6 in phase (see Fig. \ref{fig:ppc}).

\subsection{Long-term X-ray variability}

The regions of the LMC where our new BeXRBs are located were little observed in X-rays in the past. 
We used the HIgh-energy LIght curve GeneraTor \citep[HILIGT;][]{2022A&C....3800529K,2022A&C....3800531S}\footnote{\url{http://xmmuls.esac.esa.int/upperlimitserver/}} to search for serendipitous observations of the new systems with \rosat, \swift and \xmm and determine upper limits  (2$\sigma$) when the sources were not detected.
We also checked for possible detections in the \cxo source catalogue \citep{2010ApJS..189...37E} and did not find any X-ray counterpart within 10\arcsec\ of the corresponding source positions.

\subsubsection{\xmma}

Apart from the detection in the \xmm observation reported here, only upper limits are available for \sxmma from \rosat/PSPC (all-sky survey and three pointed) observations, one \swift/XRT observation and from \xmm, which slewed across the source 10 times. All the upper limits are higher than the \xmm flux we measured from the pointed observation in Jan. 2018 and only restrict the maximum 0.2--12\,keV flux to less than $\sim$8.4\ergcm{-12}. Hence, no strong outburst was detected more luminous than $\sim$2.5\,\ergs{36} and the source is likely a low-luminosity BeXRB.

\subsubsection{\xmmb}

Similarly, the position of \sxmmb was covered by the same \rosat observations. The source was detected on 1993 Oct. 14 with a 0.2--2\,keV count rate of 0.0154$\pm$0.0054\,\cts, which corresponds to a 0.2--10\,keV flux of 2.27$\pm$0.80 \ergcm{-12}, assuming the spectral model inferred from the \xmm observations and using the \rosat PSPC detector response. 
The source was also detected during one (of 12) \xmm slews (on 2006 Aug. 5) with a 0.2--12\,keV count rate of 0.37$\pm$0.19\,\cts  \citep{2008A&A...480..611S,2018yCat.9053....0X}, consistent with the fluxes measured during observations 0804550101 (0.456$\pm$0.006\,\cts) and 0841660301  (0.232$\pm$0.005\,\cts).
All detections of \xmmb indicate long-term flux variations by at most a factor of a few in a range of $\sim$1--3\,\ergcm{-12}. The corresponding luminosities between $\sim$3--9\,\ergs{35} suggest a persistent intermediate luminosity BeXRB.
No \swift observation covered the position of \sxmmb.

\subsubsection{\rxjc}

This BeXRB was frequently observed by \rosat, 26 times with the PSPC (including the all-sky survey) and 14 times with the HRI detector. 
Seven PSPC detections yielded 0.2--2\,keV count rates between 0.01\,\cts and 0.08\,\cts, while 19 upper limits are derived with similar count rates of 0.01\,\cts -- 0.07\,\cts. While the PSPC exposure times were short, typically less than 1000\,s, the HRI observations were $\sim$10 times longer. This is reflected in 8 HRI detections of \srxjc with count rates between 0.0015\,\cts and 0.0060\,\cts and upper limits between 0.0016\,\cts and 0.0065\,\cts. To take the different sensitivity of the two \rosat detectors into account we converted the count rates into 0.2--12\,keV fluxes using the detector response files and assuming the best fit \fdcut model parameters from the \xmm observation.
The inferred flux to count rate conversion factors are 1.12\ftocr{-10} and 3.09\ftocr{-10} for the PSPC and HRI detectors, respectively. The fluxes measured by \rosat are in the range of 4.6\ergcm{-13} to 8.9\ergcm{-12}, i.e. the luminosity was never observed above $\sim$2.6\,\ergs{36}, very similar to \sxmma.
No \swift observation covered the position of \srxjc.

\subsection{SALT}
\label{subsec:salt}

Optical spectroscopy of our new HMXBs was undertaken on 11 January 2022  using the High Resolution Spectrograph \citep[HRS, ][]{2014SPIE.9147E..6TC} on SALT under the transient follow-up program. The HRS is a dual-beam, fibre-fed Echelle spectrograph providing spectra in the 3700--5500\,\AA\ and 5500--8900\,\AA\ wavelength ranges. The low resolution mode (R$\sim$14 000) of the HRS was used to observe two of the targets, J0457 and J0524, with exposure times of 1200\,s. The primary reductions, which include overscan correction, bias subtraction and gain correction, were carried out with the SALT science pipeline \citep{2015ascl.soft11005C}. The remaining reduction steps which include background subtraction, arc line identification, removal of the blaze
function, and merging of orders were performed with the \textsc{midas feros} \citep{1999ASPC..188..331S} and \textsc{echelle} \citep{1992ESOC...41..177B} packages (for a detailed overview of the reduction steps, see \citealt{2016MNRAS.459.3068K}). 


\subsubsection{\xmmb}

\begin{figure}
  \centering
  \resizebox{0.95\hsize}{!}{\includegraphics[clip=]{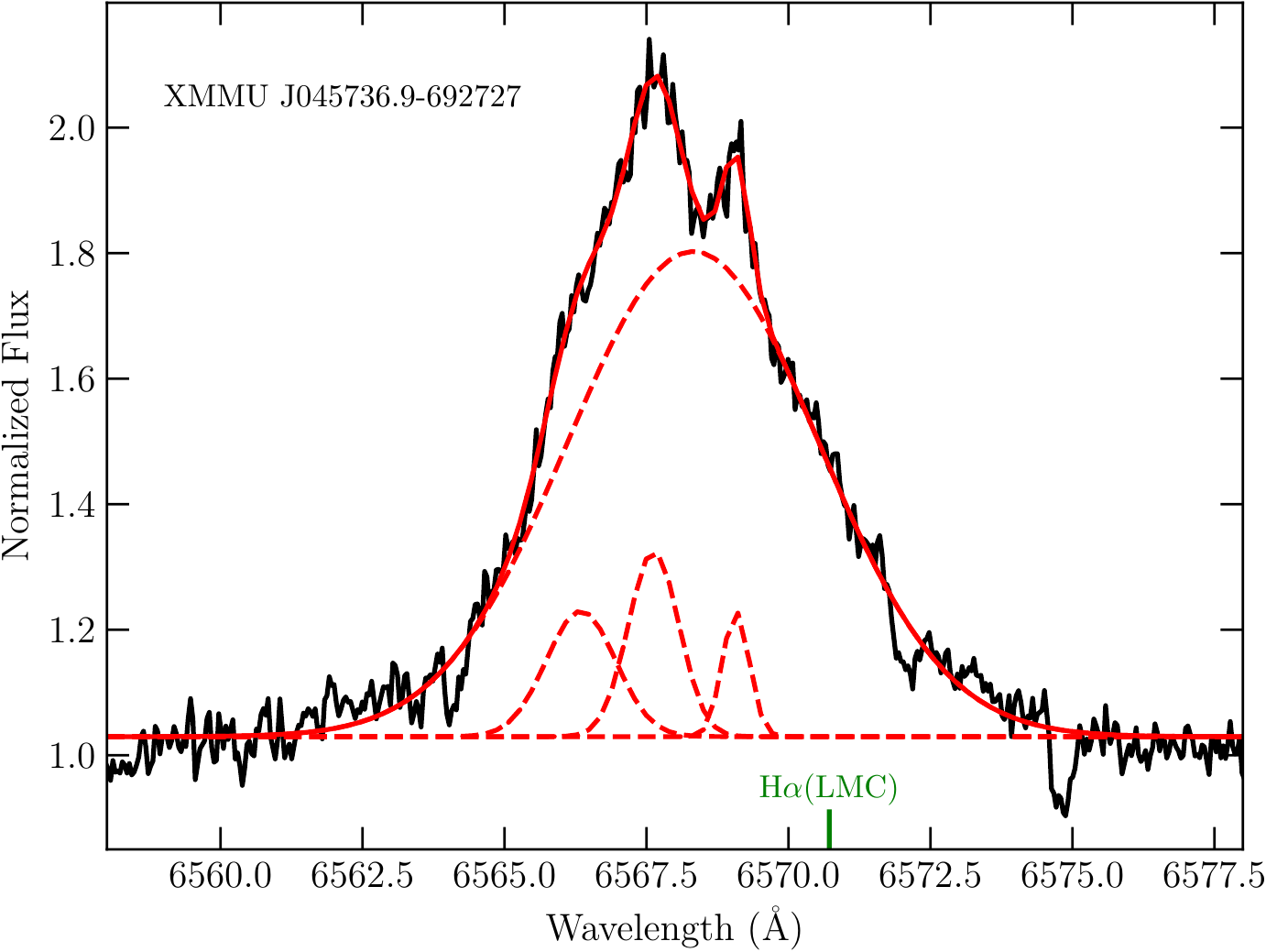}}
  \resizebox{0.95\hsize}{!}{\includegraphics[clip=]{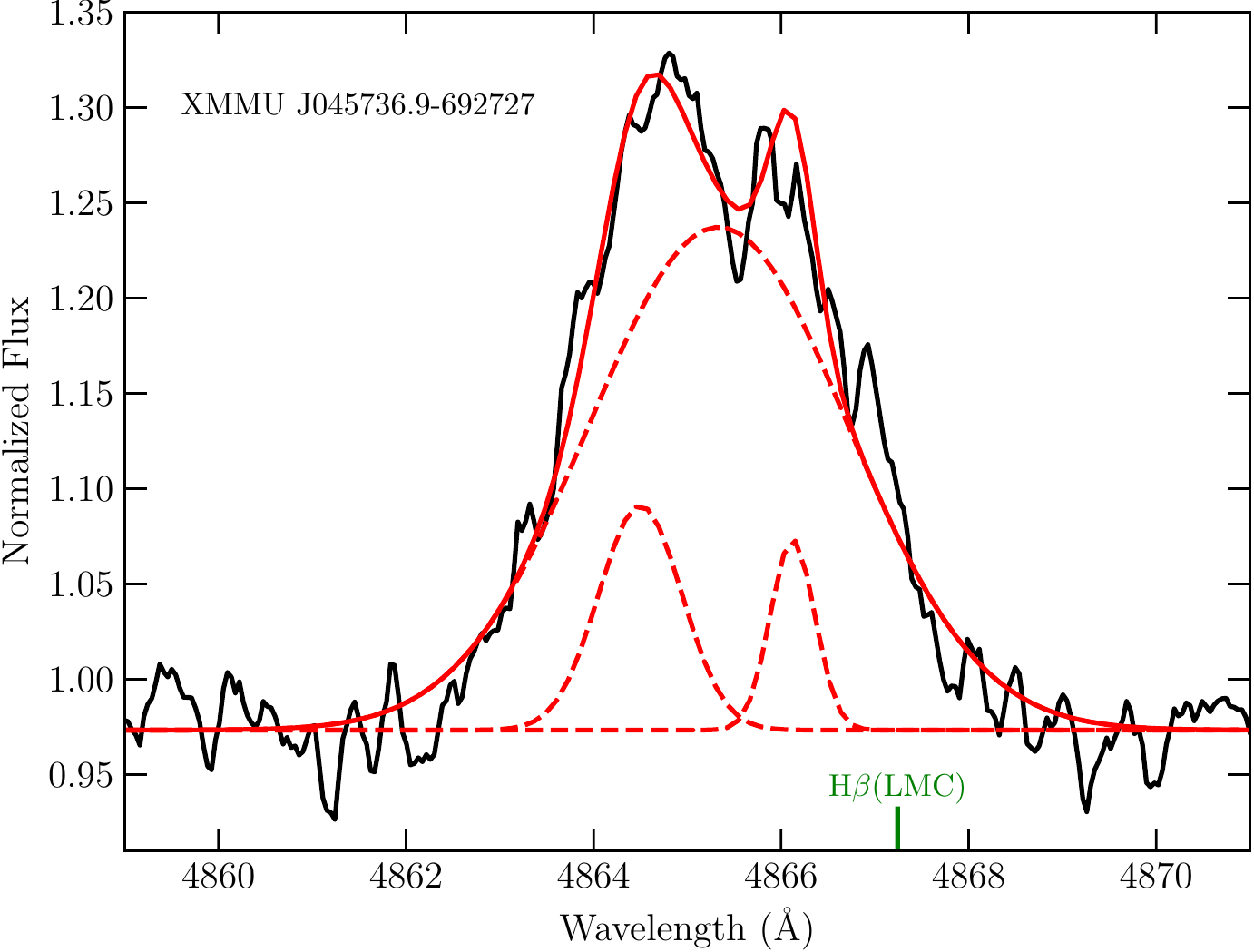}}
  \caption{
    SALT HRS spectra of \sxmmb around the \Halpha (top) and \Hbeta (bottom) lines. Both lines are in emission and highly structured.
    The spectrum around \Hbeta is smoothed using a rolling average filter with kernel size 5.
    The wavelengths shifted to the rest system of the LMC are indicated by the green vertical lines.
  }
  \label{fig:saltb}
\end{figure}

For \sxmmb we obtained the red and blue HRS spectra, which cover the \Halpha and \Hbeta lines (Fig.\,\ref{fig:saltb}).
Both lines are seen in emission (Fig.~\ref{fig:saltb}) and have measured line equivalent widths (EQWs) of -7.99$\pm$0.37\,\AA\ and -1.29$\pm$0.08\,\AA, respectively.
Their complex morphology is characterised by a  blue-dominated double-peaked structure.
Shoulders on both sides of the line blend suggests more than two emission components.
We have modelled the line profiles with a superposition of several Gaussian lines. 

For the \Halpha line four Gaussians are required to obtain an acceptable fit, 
their central wavelengths and the widths are detailed in Table\,\ref{tab:hrs}.
The \Halpha line complex is dominated by a broad line, while narrower lines are responsible for the two peaks.
An additional narrow line is contributing on the blue side.
We have verified that the faint lines are not caused by an imperfect sky-subtraction. Inspecting the sky-background spectrum shows a weak narrow \Halpha line, but at a different wavelength, which is properly subtracted.
The profile of \Hbeta is very similar to that of \Halpha and may be composed by the same number of lines.
However, due to the lower signal three lines are formally sufficient to reproduce the profile (Fig.~\ref{fig:saltb}).
Again a broad line dominates and two narrower lines are responsible for the two peaks. 

For the case of \Halpha, converting the relative wavelength shifts into velocities along the line of sight, 
the narrower lines correspond to -90.4\,\kms, -31.8\,\kms and 34.3\,\kms, relative to the centre of the broad line.
The dominant broad line is blue-shifted with respect to the LMC rest frame (assuming z$_{\rm LMC}$ = 0.00093) by -109.3\,\kms.

\subsubsection{\rxjc}

\begin{figure}
  \centering
  \resizebox{0.95\hsize}{!}{\includegraphics[clip=]{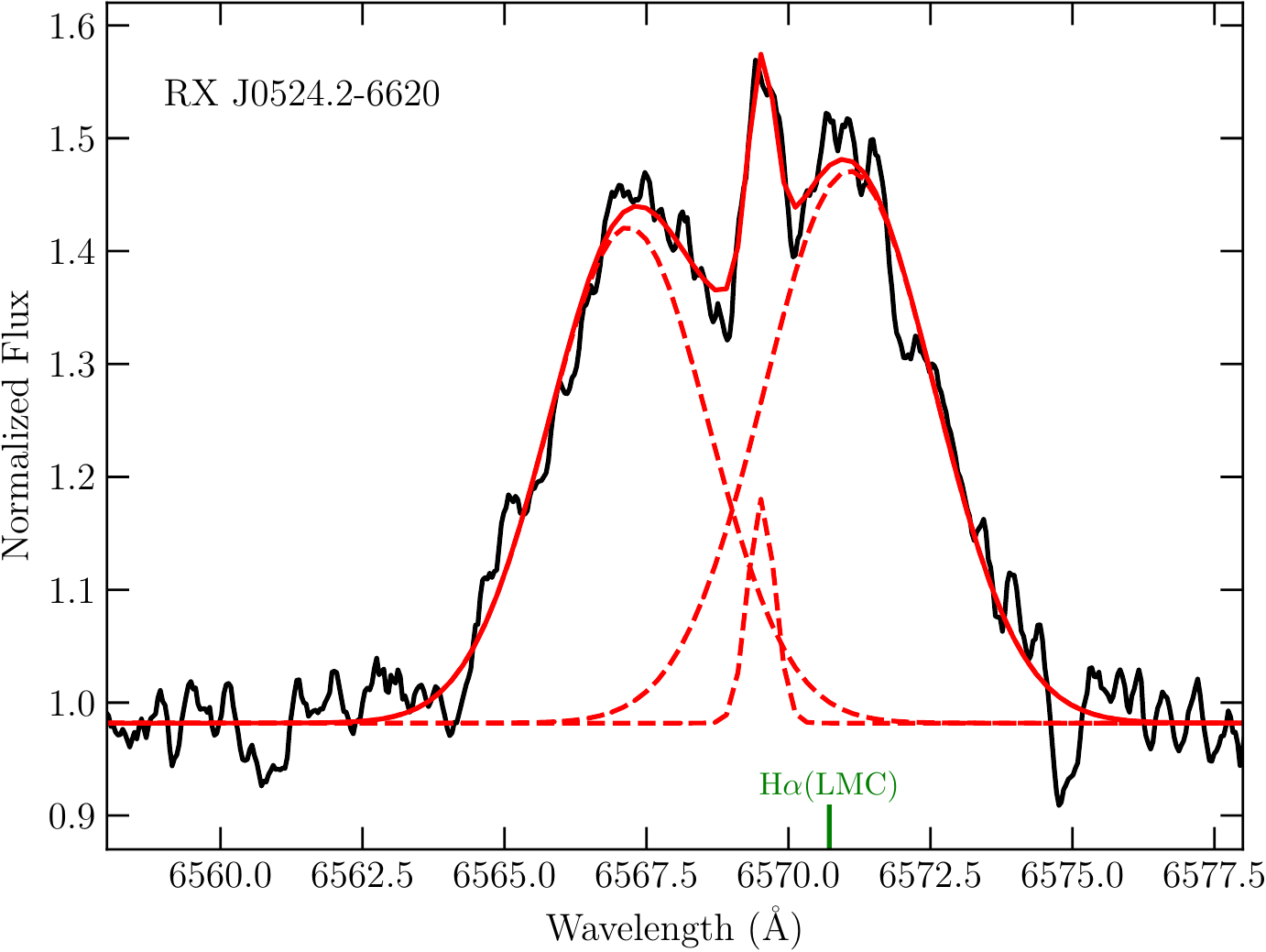}}
  \caption{
    SALT HRS spectrum of \srxjc around the \Halpha line, marked as in Fig.\,\ref{fig:saltb}. The narrow line is a residual feature from 
    sky background subtraction.
    The spectrum is smoothed using a rolling average filter with kernel size 5.
  }
  \label{fig:saltc}
\end{figure}

The \Halpha line from \srxjc is seen in emission and has a triple-peak structure (Fig.~\ref{fig:saltc}), with a measured EQW of -4.20$\pm$0.24\,\AA.
We fitted the line profile with three Gauss functions and again list their central wavelengths and widths in Table\,\ref{tab:hrs}. 
The blue and red components have similar widths (1.4--1.5\,\AA) while the central line is significantly narrower (0.24\,\AA). Inspection of the sky background spectrum reveals a strong narrow \Halpha line, which leaves some residual after subtraction.
The blue and red components are shifted with respect to each other, corresponding to a relative line of sight velocity of -177.6\,\kms. 

\begin{table}
\centering
\caption[]{Emission line profiles.}
\label{tab:hrs}
\begingroup
\begin{tabular}{cclc}
\hline\hline\noalign{\smallskip}
\multicolumn{1}{c}{Source} &
\multicolumn{1}{c}{Line} &
\multicolumn{1}{c}{Center} &
\multicolumn{1}{c}{$\sigma$ width} \\
\multicolumn{1}{c}{name} &
\multicolumn{1}{c}{} &
\multicolumn{1}{c}{(\AA)} &
\multicolumn{1}{c}{(\AA)} \\

\noalign{\smallskip}\hline\noalign{\smallskip}
\sxmmb & \Halpha & 6566.343 $\pm$ 0.251 & 0.608 $\pm$ 0.216 \\
       &         & 6567.626 $\pm$ 0.103 & 0.445 $\pm$ 0.088 \\
       &         & 6568.323 $\pm$ 0.064	& 2.213 $\pm$ 0.061 \\
       &         & 6569.074 $\pm$ 0.036	& 0.237 $\pm$ 0.047 \\
\noalign{\smallskip}
\sxmmb & \Hbeta  & 4864.499 $\pm$ 0.060 & 0.445 fixed       \\
       &         & 4865.333 $\pm$ 0.048 & 1.382 $\pm$ 0.042 \\
       &         & 4866.123 $\pm$ 0.072 & 0.237 fixed       \\
\noalign{\smallskip}
\srxjc & \Halpha & 6567.182 $\pm$ 0.086 & 1.405 $\pm$ 0.076 \\
       &         & 6569.523 $\pm$ 0.036\tablefootmark{a} & 0.239 $\pm$ 0.034 \\
       &         & 6571.075 $\pm$ 0.094 & 1.494 $\pm$ 0.079 \\
\noalign{\smallskip}\hline
\end{tabular}
\endgroup
\tablefoot{
Errors are given for a 90\% confidence range.
\tablefoottext{a}{Residual feature from sky background subtraction.}
}
\end{table}

\subsection{OGLE}
\label{subsec:ogle}

The regions around the three X-ray sources were monitored regularly in the I and V filter bands during phases III and IV of the OGLE project. 
Images were taken in I and less frequently in V, with the photometric magnitudes calibrated to the standard VI system.
OGLE-III and OGLE-IV I-band images were reduced together using the
same reference set yielding a homogeneous OGLE-III + OGLE-IV time
series.

\subsubsection{\xmma}

\begin{figure*}[t]
  \centering
  \resizebox{\hsize}{!}{\includegraphics[clip=]{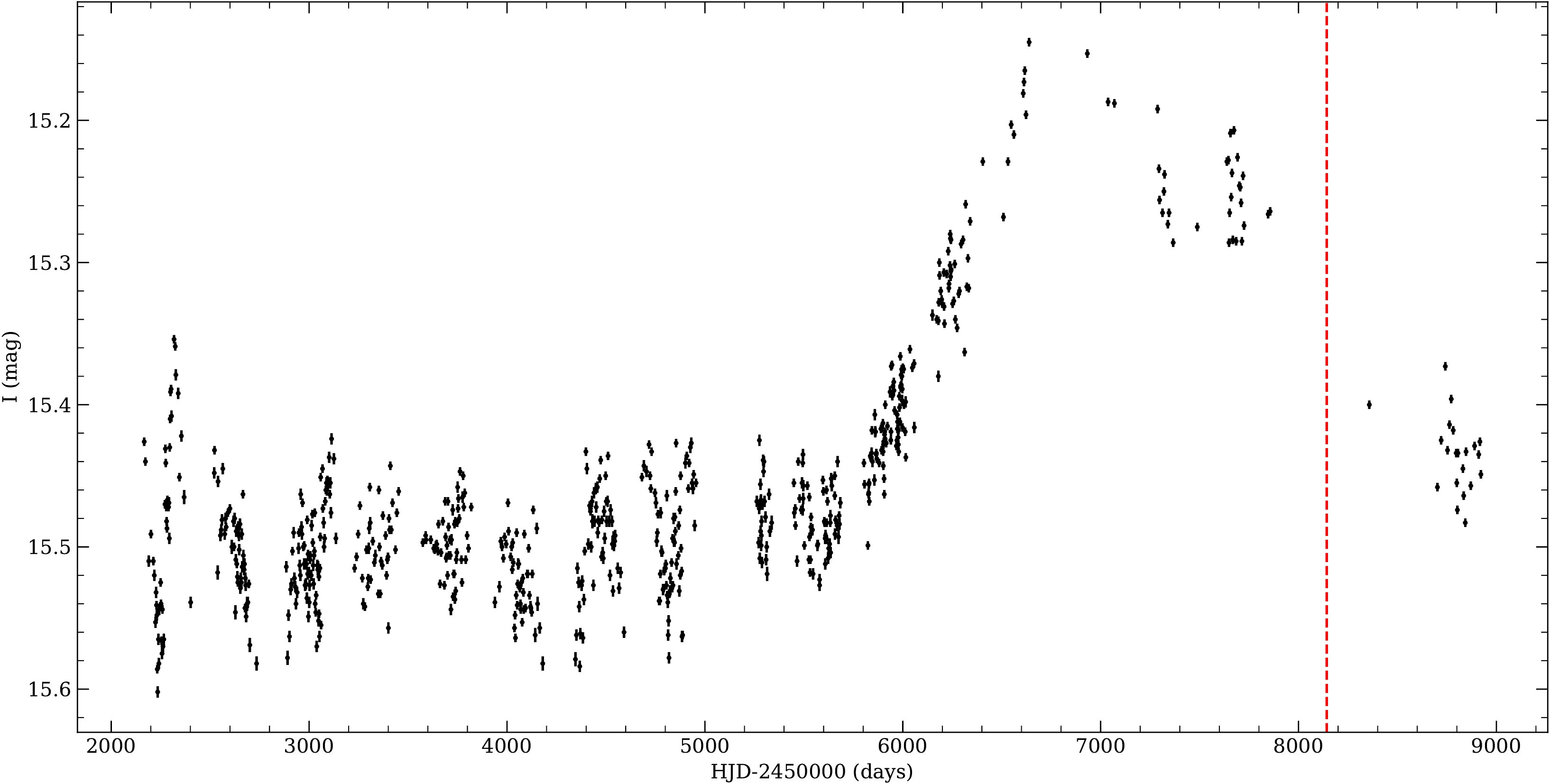}}
  \caption{
    OGLE I-band light curve of \sxmma between 2001 Sep. 14 and 2020 Mar. 13. 
    The vertical dashed red line indicates the time of the \xmm observation.
  }
  \label{fig:oglelca}
\end{figure*}

From the optical counterpart of \sxmma (2MASS\,\massa, OGLE IDs LMC135.1.36 and LMC136.4.180 from two overlapping OGLE-III fields, and LMC531.12.31 from OGLE-IV) 18.5 years of OGLE I-band data are available.
The first part of the light curve, which is presented in Fig.~\ref{fig:oglelca}, 
is characterised by brightness variations with 0.15--0.25\,mag amplitude.
However, after mid March 2011 a sudden increase in brightness by $\sim$0.3\,mag over a period of about 2.75\,years occurred, 
followed by a more gradual decrease for at least 6.25\,years. 
After this long optical outburst the brightness reached almost the level the star had before, 
when the observations ended in March 2020.

\begin{figure}
  \centering
  \resizebox{\hsize}{!}{\includegraphics[clip=]{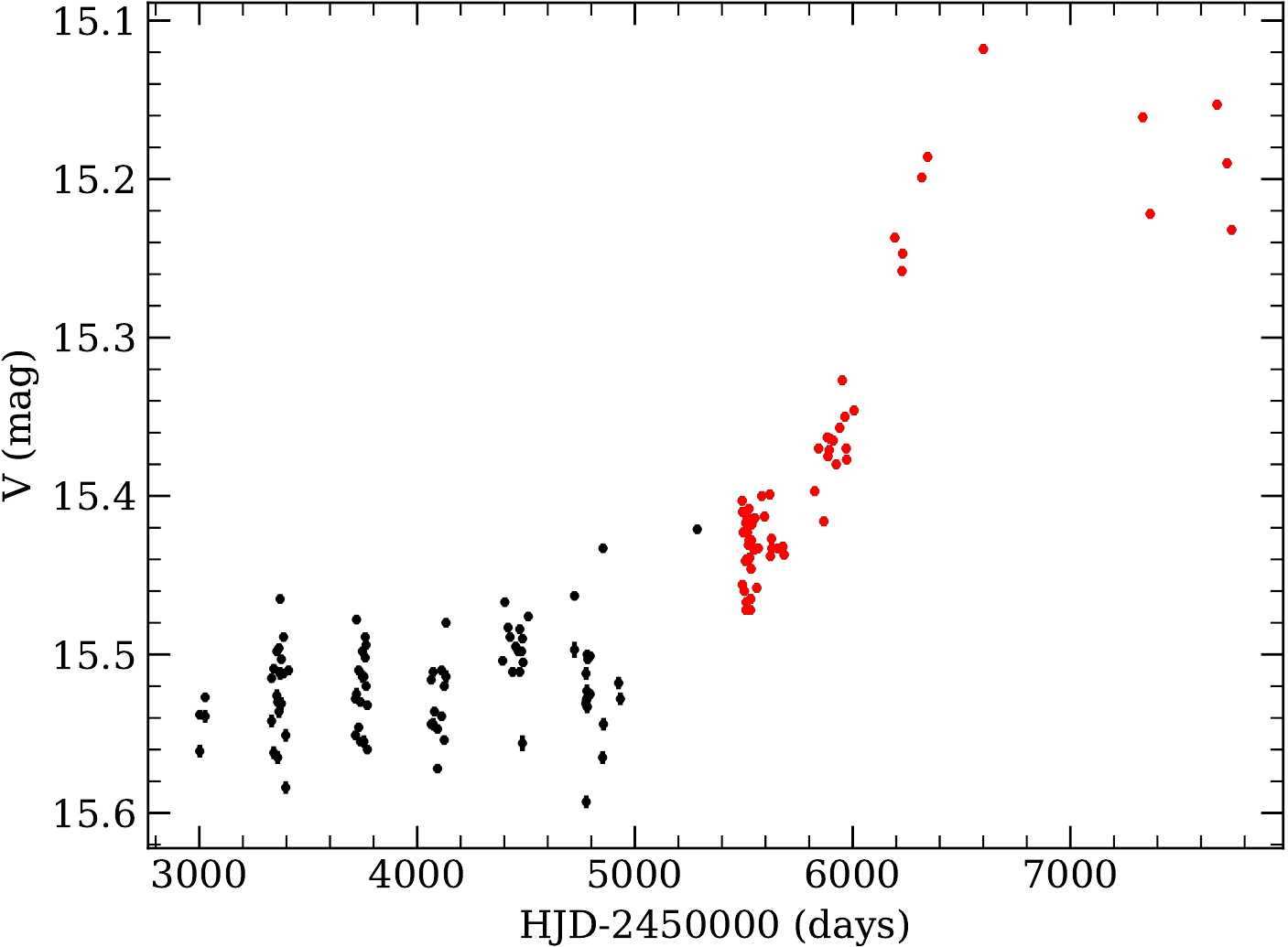}}
  \caption{
    OGLE V-band light curve of \sxmma between 2003 Dec. 28 and 2016 Dec. 18. The outburst is marked in red.
    The \xmm observation was performed after the last V-band observations. See also Fig.\,\ref{fig:oglelca}.
  }
  \label{fig:oglevlca}
\end{figure}

\begin{figure}
  \centering
  \resizebox{\hsize}{!}{\includegraphics[clip=]{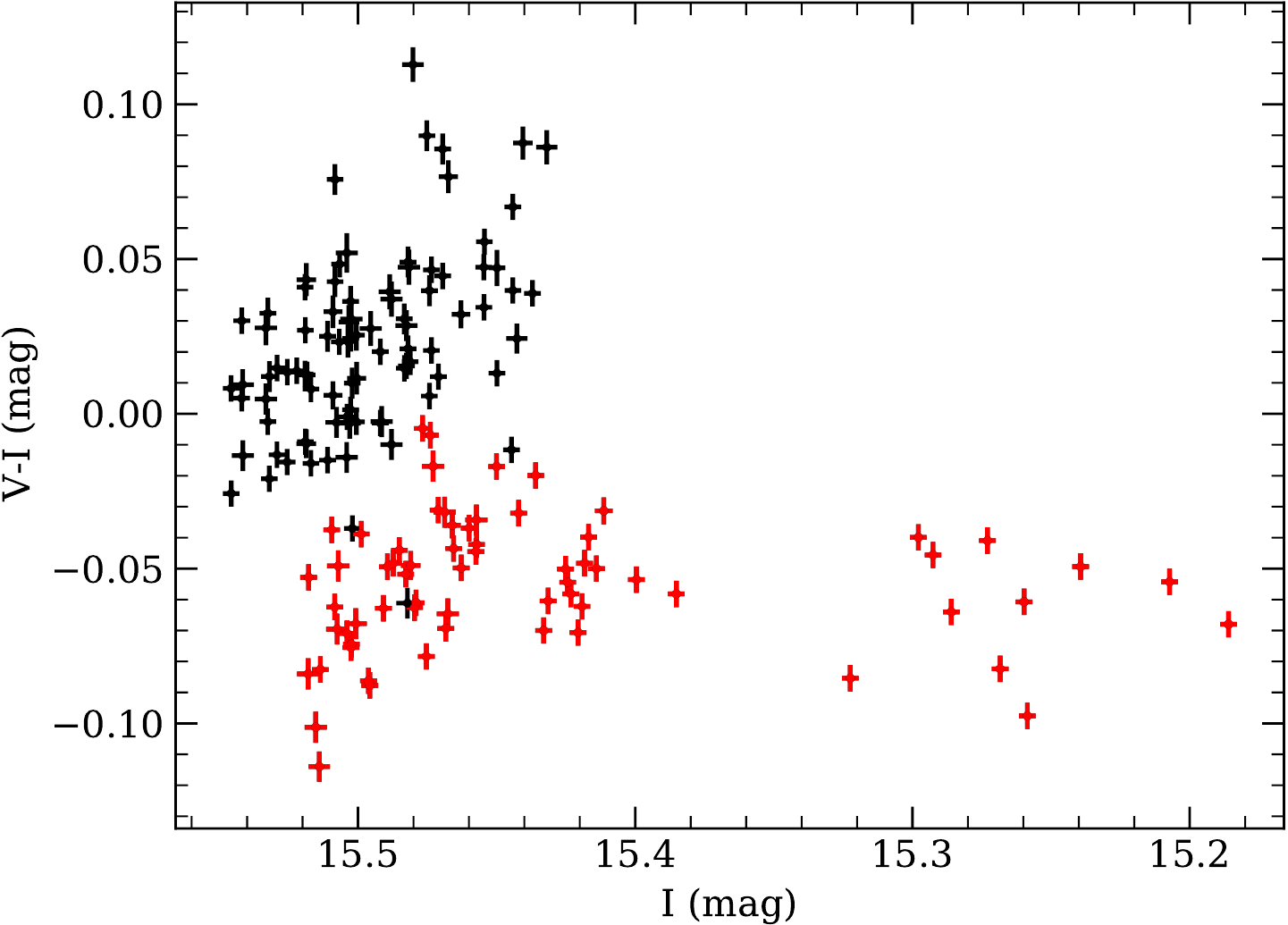}}
  \caption{
    \sxmma: OGLE V-I colour index vs. I (interpolated linearly to the times of the V-band measurements). As in Fig.\,\ref{fig:oglevlca}, the data taken during the outburst are marked in red.
  }
  \label{fig:oglecola}
\end{figure}

OGLE V-band observations of \sxmma started more than two years later than OGLE I and provided 17 years and 2 months of monitoring (Fig.~\ref{fig:oglevlca}).
To estimate the I-band magnitudes at the times of the V-band measurements we interpolated the adjacent I-band magnitudes and computed the V$-$I colour index. 
The colour index is shown as a function of the I magnitude in Fig.~\ref{fig:oglecola}. 
Data taken during the optical outburst are marked in red in Figs.~\ref{fig:oglevlca} and \ref{fig:oglecola}. During the outburst the BeXRB appears bluer than during the optical low state, but independent of brightness.


\begin{figure}
  \centering
  \resizebox{0.98\hsize}{!}{\includegraphics[clip=]{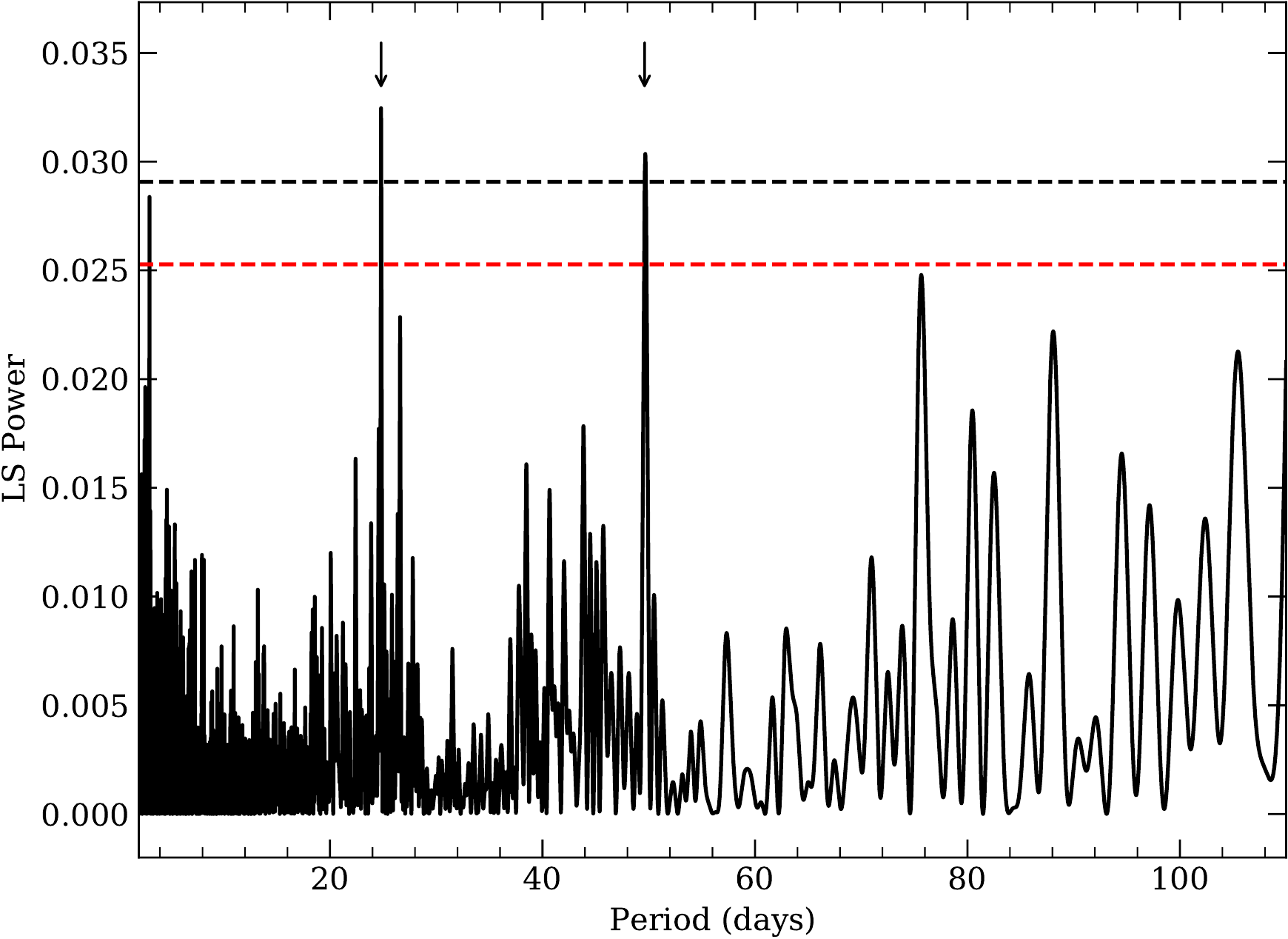}}
  \caption{
    LS periodogram of the OGLE I-band data (before the large outburst, HJD $<$ 2456000 days) of \sxmma.
    The arrows mark the highest peaks at 49.61\,days and 24.81\,days. The red and black dashed lines mark the 95\% and 99\% confidence levels.
  }
  \label{fig:oglelsa}
\end{figure}

\begin{figure}
  \centering
  \resizebox{0.98\hsize}{!}{\includegraphics[clip=]{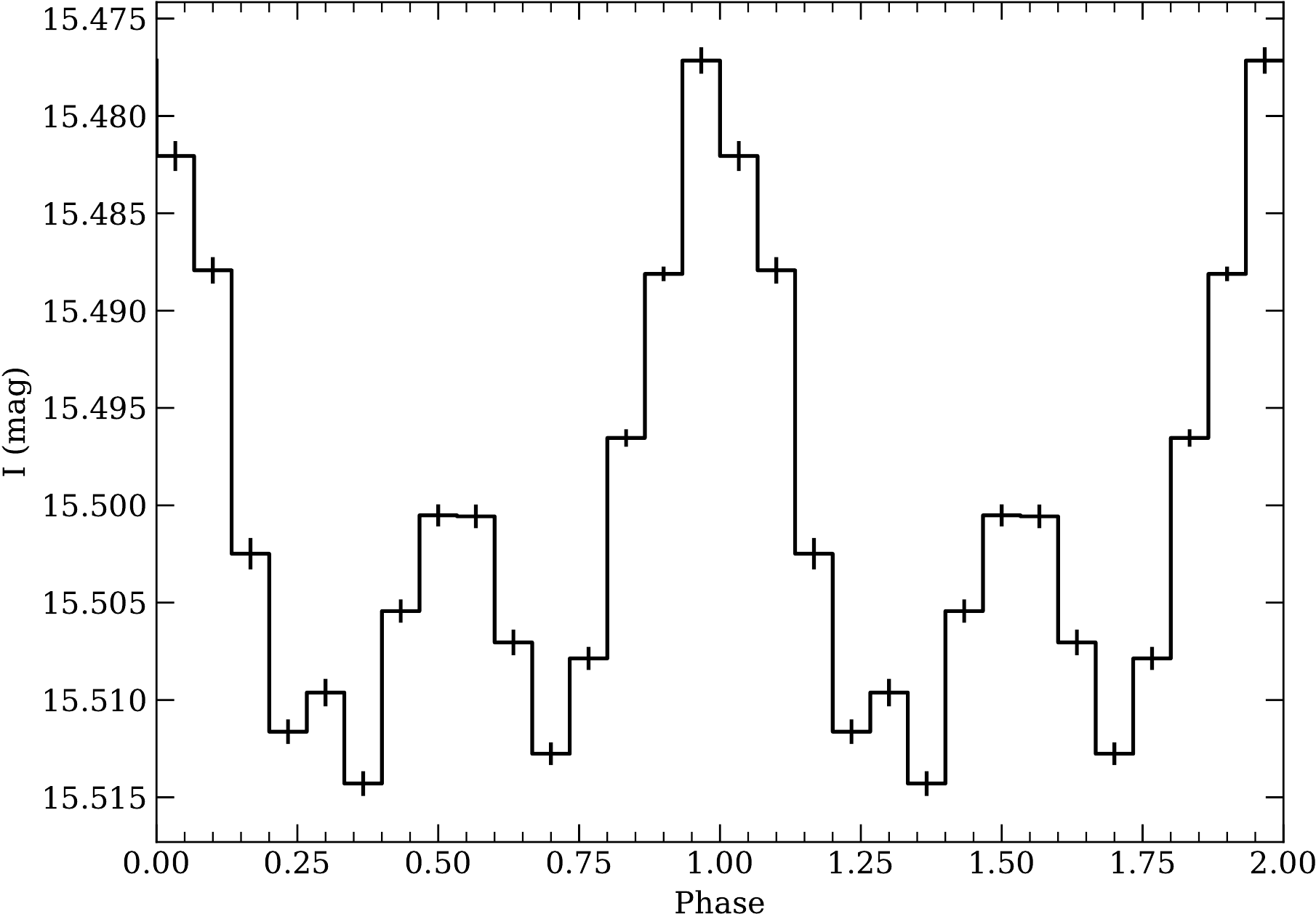}}
  \caption{
    Folded OGLE I-band light curve of \sxmma with a period of 49.61\,days, repeated for two cycles for clarity. 
  }
  \label{fig:oglefolda}
\end{figure}

We searched for periodic variations in the OGLE I-band light curve. To avoid the large brightness variations during the long outburst, we selected the data before 
HJD of 2456000\,days and used again the LS periodogram analysis for periods between 2 and 110\,days. 
The LS periodogram revealed two highly significant peaks at periods of 24.81$\pm$0.05\,days and 49.61$\pm$0.20\,days (see Fig.\,\ref{fig:oglelsa}). The errors are conservatively estimated from the width of the peaks. The longer period is consistent with twice the value of the shorter one, which suggests 24.81\,days as first harmonic of a fundamental period of 49.61\,days. 
Fig.\,\ref{fig:oglefolda} shows the light curve (before the outburst) folded with the period of 49.61\,days. It is dominated by a symmetric outburst which lasts for about half the period and also reveals a smaller outburst, shifted by 0.5 in phase, which is responsible for the harmonic signal in the LS periodogram.


\subsubsection{\xmmb}

\begin{figure*}[t]
  \centering
  \resizebox{\hsize}{!}{\includegraphics[clip=]{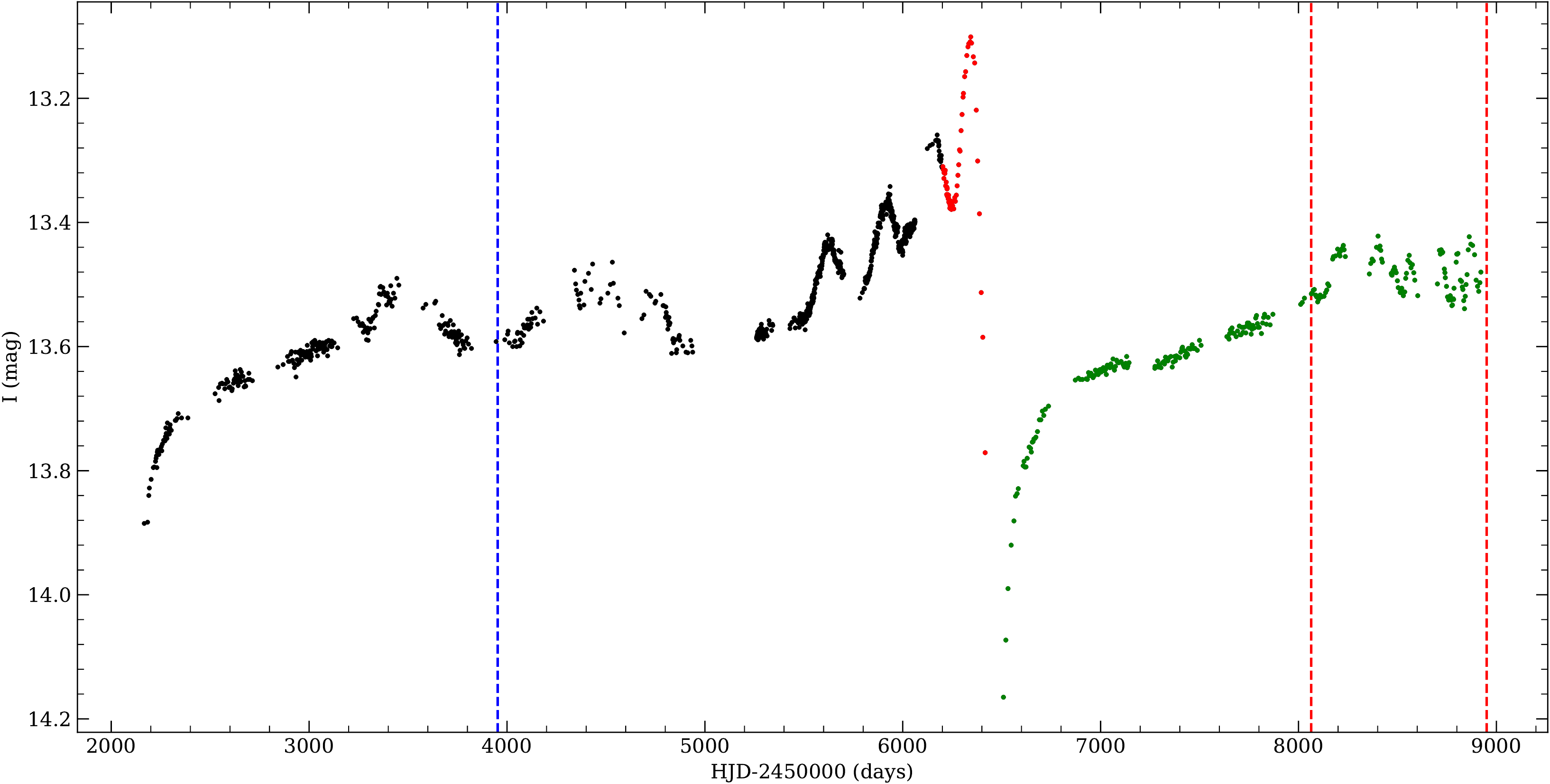}}
  \caption{
    OGLE I-band light curve of \sxmmb between 2001 Sep. 14 and 2020 Mar. 13. 
    The vertical dashed lines indicate the times of the \xmm slew detection (2006 Aug. 5; blue) and the pointed observations (0804550101 and 0841660301, Table\,\ref{tab:xmmobs}; red).
  }
  \label{fig:oglelcb}
\end{figure*}

The OGLE I-band light curve of the most likely counterpart of \sxmmb (2MASS\,\massb, OGLE IDs LMC127.8.34098 and LMC531.09.34322 from OGLE-III and OGLE-IV, respectively) over 18.5 years is presented in Fig.\,\ref{fig:oglelcb}. 
Several outbursts are visible, in particular a very strong one, which is followed by a dramatic decrease in brightness by about one magnitude. A similar event was likely largely missed at the beginning of the OGLE monitoring where only part of the recovery from a dip was recorded. The two events were $\sim$4395\,days apart.

\begin{figure}
  \centering
  \resizebox{\hsize}{!}{\includegraphics[clip=]{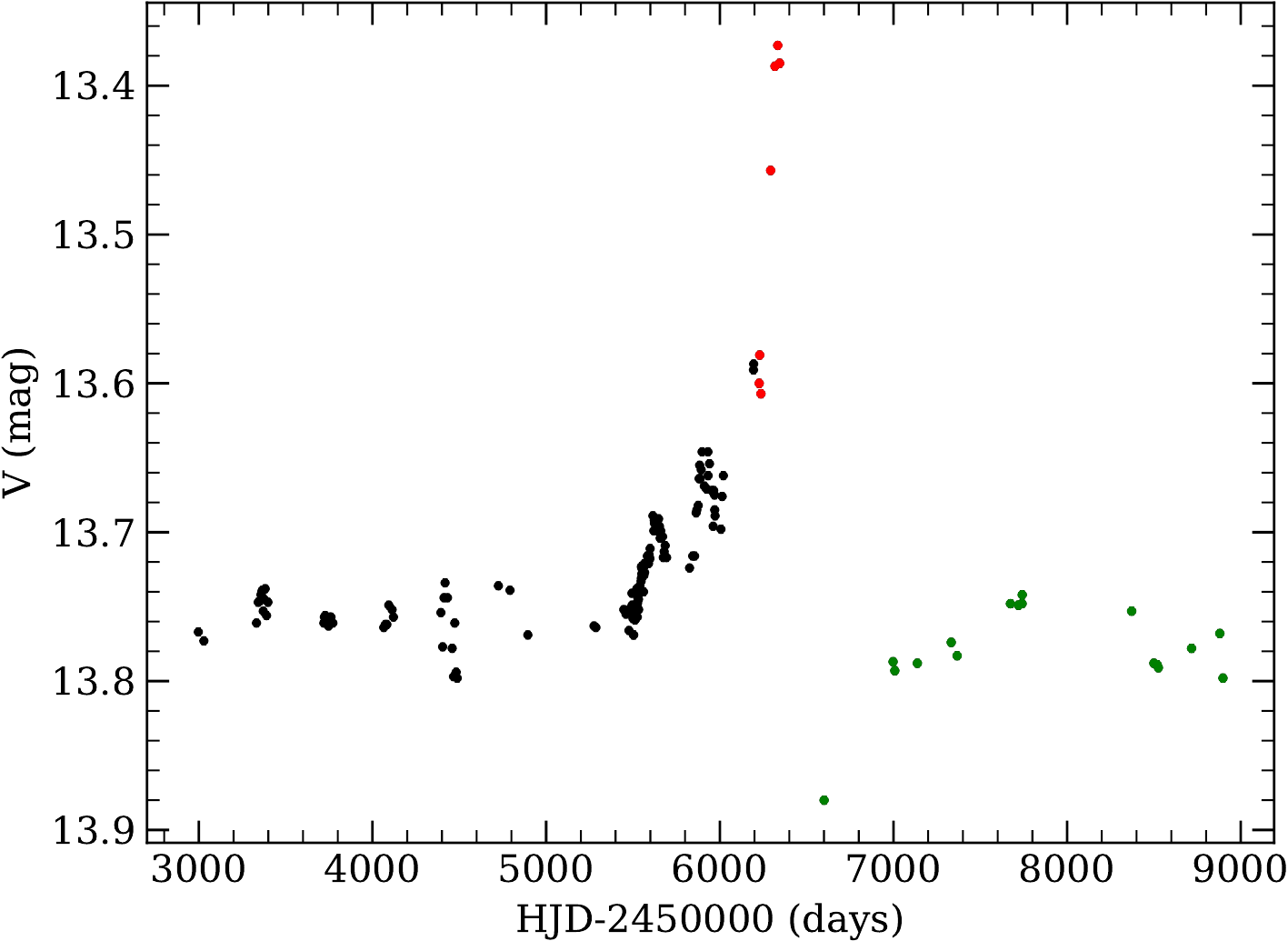}}
  \caption{
    OGLE V-band light curve of \sxmmb between 2003 Dec. 23 and 2020 Feb. 18. 
    See also Fig.\,\ref{fig:oglelcb}.
  }
  \label{fig:oglevlcb}
\end{figure}

\begin{figure}
\centering
  \resizebox{\hsize}{!}{\includegraphics[clip=]{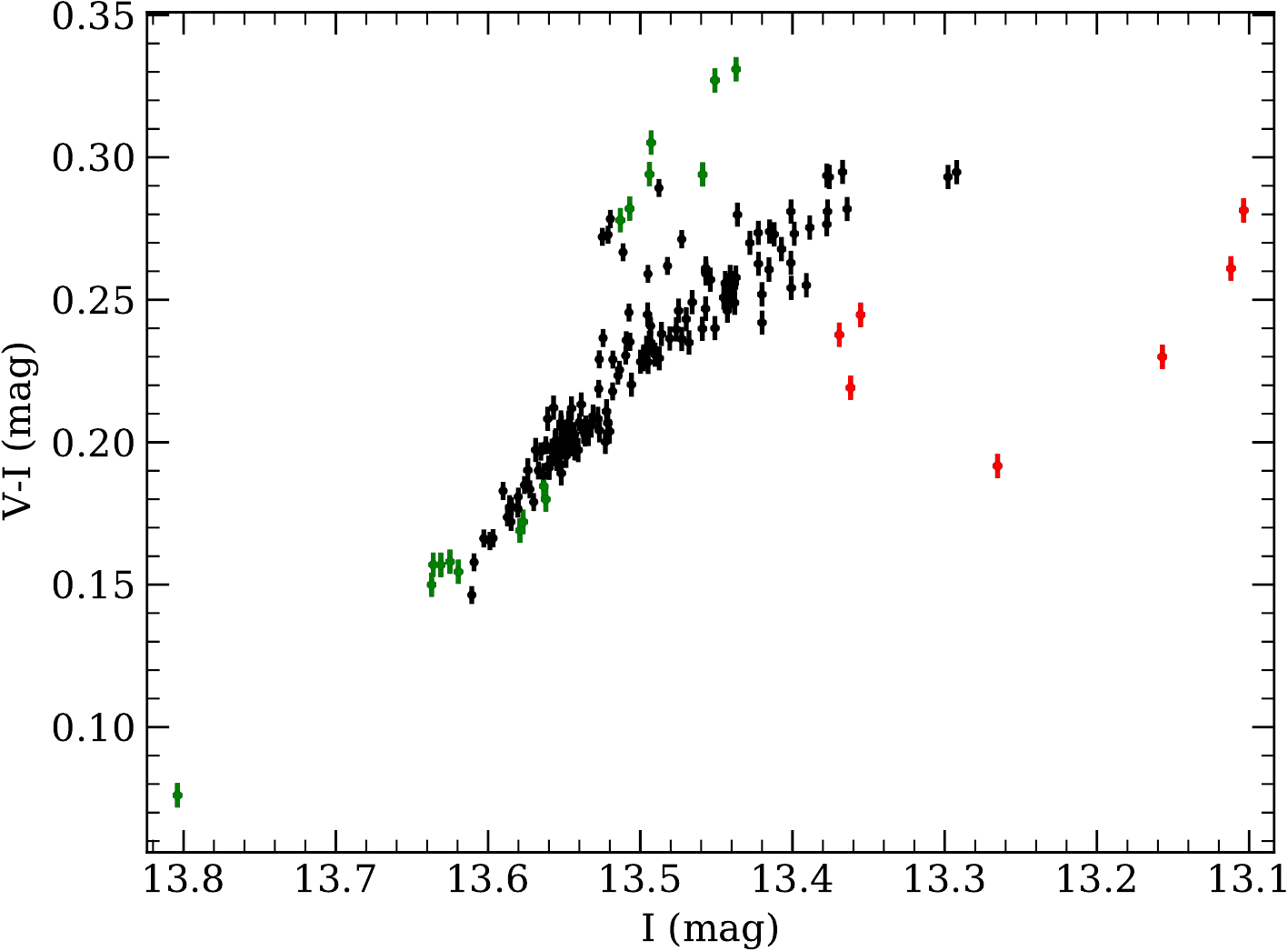}}
  \caption{
    \sxmmb: OGLE V-I colour index vs. I (interpolated linearly to the times of the V-band measurements). 
  }
  \label{fig:oglecolb}
\end{figure}

\sxmmb was also monitored in the V-band between end of 2003 and February 2020.
Unfortunately, during the deep dip only one V-band measurement is available (Fig.\,\ref{fig:oglevlcb}).
The V-I colour index as function of I is presented in Fig.\,\ref{fig:oglecolb}.
Before the major outburst, the system is characterised by reddening when becoming brighter (black data points in Figs.\,\ref{fig:oglelcb} to \,\ref{fig:oglecolb}).
However, during the major outburst, V-I did not increase further (red data points). During the rising part of the light curve after the deep dip, V-I followed the relation from the first part of the light curve and reached maximum values when the pre-outburst level was recovered (green). 

The three outbursts before the giant one indicate a regular pattern, which might be related to the orbital period of the BeXRB system.
To remove long-term trends, following \citet{2019MNRAS.490.5494M}, we first subtracted a smoothed light curve 
\citep[derived by applying a Savitzky–Golay filter with a window length of 401 data points;][]{1964AnaCh..36.1627S} from the OGLE I-band light curve.
To search for periodic variations, we again applied the LS periodogram analysis. 
We first used I-band data between 13.2\,mag and 13.6\,mag to clip the largest variations. The LS periodogram revealed a period of 288\,days.
However, when marking the times of expected outburst maxima in the light curve, no other peak with the right phase can be recognised. Alternatively, we selected only times between HJD 2455000 and  2456250 (the three strong outbursts) and a period of 273.5\,days was found. This fits the three outbursts very well, but again no other peak in phase could be identified.


\subsubsection{\rxjc}

The position of the most probable counterpart of \srxjc, 2MASS\,\massc was monitored during the OGLE-IV survey in two overlapping fields. For the I-band light curve we combined data sets LMC505.26.12019 and LMC519.16.21009, apart from the standard calibration we identified that photometric magnitudes from the two fields differ by about 0.03 mag, which we subtracted from the second data-set in order to match photometric values.
For the V-band light curve we combined the data sets LMC505.26.16914 and LMC519.16.20240 while a correction of 0.027 mag was added.

\begin{figure*}[t]
  \centering
  \resizebox{\hsize}{!}{\includegraphics[clip=]{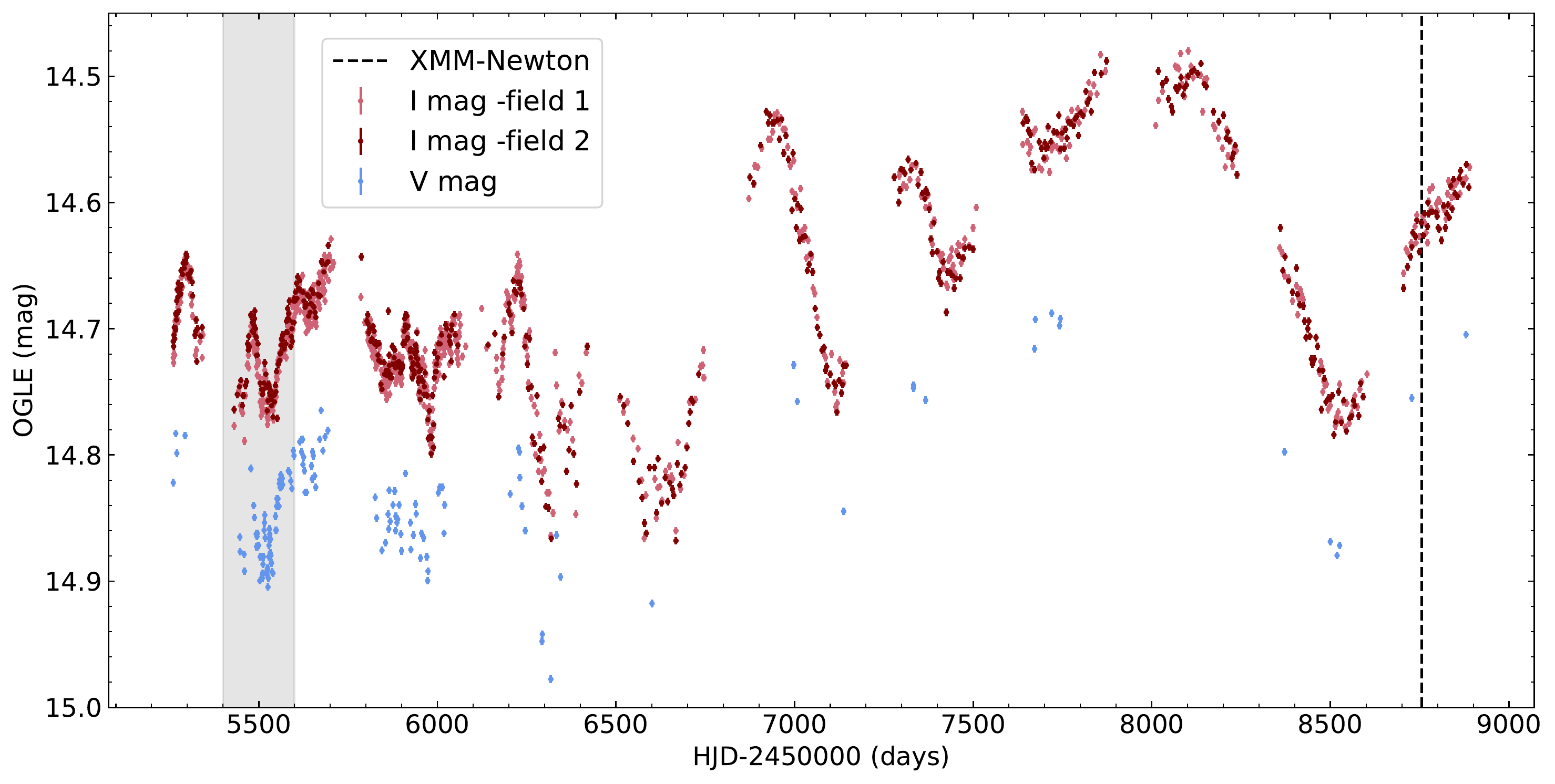}}
  \caption{
    OGLE I and V band light curves of \srxjc. 
    The vertical dashed red line indicates the time of the \xmm observation. The shaded region marks the period with more dense monitoring in the V filter and can be used to track colour evolution through a ``flare''.
  }
  \label{fig:oglelcc}
\end{figure*}

The I band is characterised by ``outbursts'' with patterns that resemble a quasi-periodic behaviour (see Fig. \ref{fig:oglelcc}). The early light curve (MJD 55200--56500) shows several short flares, with duration of about or less than 100 days. At later times (MJD 56500--59000), the light curve is characterised by four larger events with year-long rise and fall times. To investigate the presence of a periodic signal we computed the LS periodogram for the complete light curve, but also for two separate intervals, the early part (i.e. before MJD 56800) and the later part. The periodogram revealed significant peaks consistent with quasi-periodic behaviour and in agreement with the flaring periods. Otherwise, apart from some peaks around 1 day, no other periodic signal was evident.

\begin{figure}
  \centering
  \resizebox{\hsize}{!}{\includegraphics[clip=]{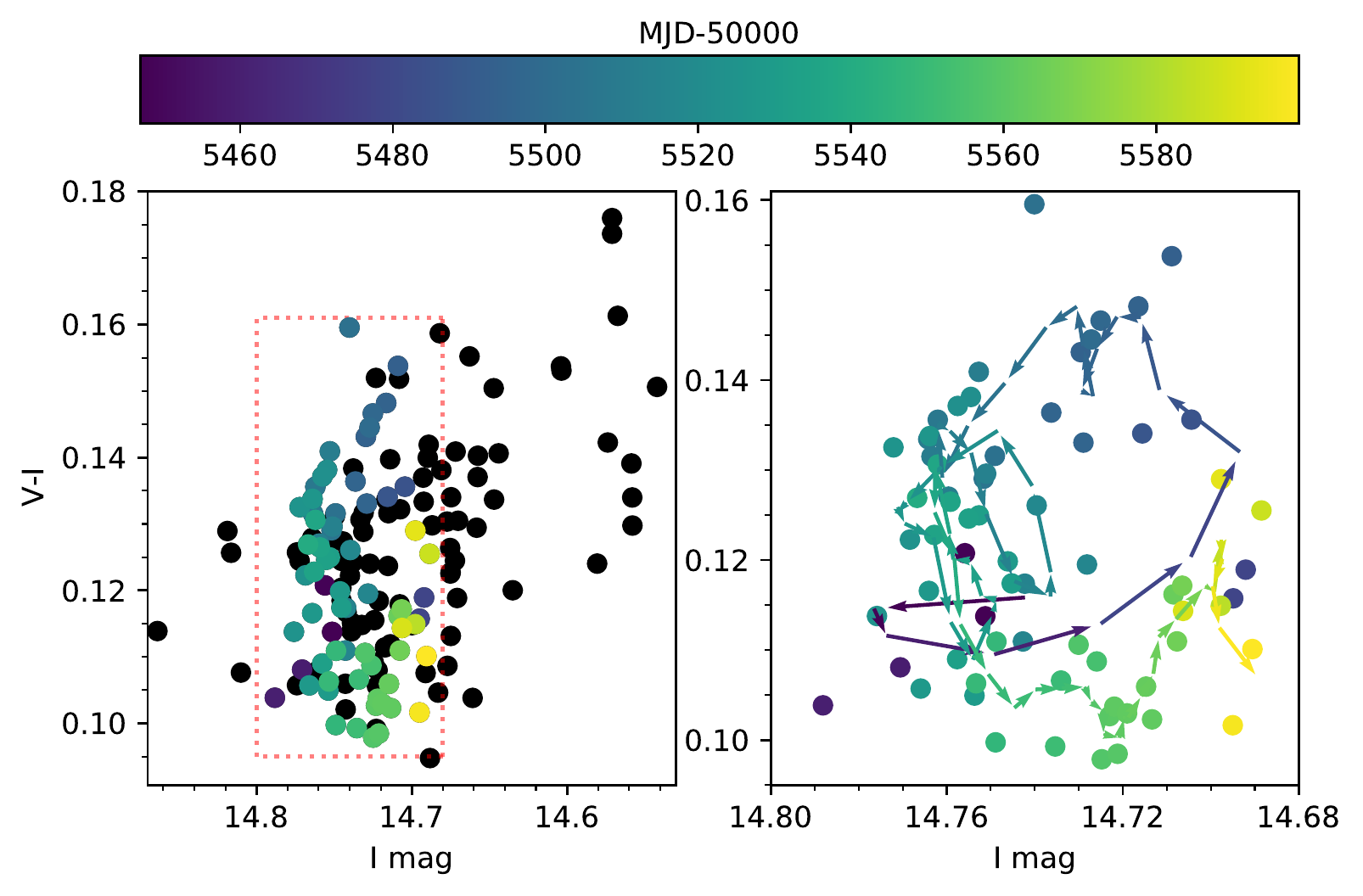}}
  \caption{
    \srxjc: OGLE V-I colour index vs. I (interpolated linearly to the times of the V-band measurements). In the left panel we plot the complete data set, while in the right panel we restrict the times to MJD 55400--55600 (see shaded region in Fig.\,\ref{fig:oglelcc}). Apart from a general trend of redder when brighter, smoothing the date reveals a hysteresis loop (see right panel) where the rise of a flare is bluer than the decay.
  }
  \label{fig:oglecolc}
\end{figure}

Similarly to the other two systems, we interpolated the I-band magnitudes and computed the V$-$I colour index (see Fig. \ref{fig:oglecolc}). 
There appears to be a general trend of being redder when brighter.
Moreover, when focusing on the earlier OGLE epochs and in the range of MJD 55400--55600, where OGLE covered a full ``flare'', we studied this evolution in more detail. By doing so we find a hysteresis loop, showing that as the source becomes brighter it also becomes redder, however when the maximum is reached the decay track is much redder than the rise for the same I-band magnitude.

\section{Discussion}
\label{sec:discussion}

From a search for new HMXBs in our \xmm observations of supernova remnant candidates in the supergiant shells LMC5 and LMC7 in the north and west of the LMC, we report the discovery of three new BeXRBs. Two of them show X-ray pulsations with 317.7\,s and 360.7\,s.

\subsection{\xmma}
\label{subsec:discua}

\sxmma was detected in an \xmm observation from Jan. 2018. 
The X-ray spectrum is of low statistical quality and can be modelled with an absorbed power law with a photon index of $\sim$1.2, consistent with values reported from other BeXRBs in the Magellanic Clouds, but on the steep end of the distribution \citep[90\% of the SMC systems show indices between 0.74 and 1.26,][]{2004A&A...414..667H}. The inferred X-ray luminosity of 6.5\ergs{34} suggests a low-luminosity BeXRB without any detected outburst brighter than $\sim$2.5\,\ergs{36}.

We identify a V=15.5\,mag star as optical counterpart which exhibits the typical brightness and colours of a Be star in the LMC. 
Photometric data obtained by OGLE in the I-band (Fig.\ref{fig:oglelca}) reveals periodic variations around a mean value of $\sim$15.6\,mag, at least for the first 9.5\,years. 
An LS periodogram of the I-band light curve from this epoch (Fig.\ref{fig:oglelsa}) shows two significant peaks at periods of 49.6\,days and 24.8\,days, the shorter being the first harmonic of the longer (fundamental) period. We suggest the 49.6\,days as orbital period of the binary system, but cannot rule out 24.8\,days completely.
If the orbital period is 24.8\,days, regularly every second orbit the optical outburst must be weak or often absent, as can be seen from Fig.\,\ref{fig:oglefolda}, where the I-band light curve is folded with 49.6\,days.
On the other hand an orbital period of 49.6\,days is also remarkable as it shows two outbursts per orbit, about 0.5 in phase apart, with one considerably brighter than the other (Fig.\,\ref{fig:oglefolda}).
To our knowledge, two optical outbursts per binary orbit have not been reported from other BeXRBs
However, an NS in a nearly circular orbit, which is tilted relative to the Be disk, approaches the disk twice about 0.5 in phase apart. Disk material may be pushed away and towards us during the two encounters that could be partly be hidden by the disk and naturally lead to two different optical outbursts.
A similar, but in some respects different phenomenon was reported by \citet{2022MNRAS.tmp..192M} for the BeXRB pulsar SXP\,15.3 in the SMC. Also for that system two periodicities, 74.5\,d and 148\,d, were found from optical (OGLE) and X-rays (\swift). However, the difference is that the optical outbursts occur regularly, but X-ray outbursts only for every second optical outburst. The authors discuss this similarly with a misalignment between the Be disk and the orbital plane of the NS, which is in a nearly circular orbit.
No clear X-ray outbursts are known from \sxmma, which suggests that the NS does not approach the Be disk as close as in the case of SXP\,15.3, but still sufficiently near to disturb it. 

The V-I colour variations during the first part of the light curves (Fig.\,\ref{fig:oglecola}) are weakly correlated with brightness (I). A marginal trend of reddening with brightness is visible. When the source became brighter than I=15.42\,mag, the V-I index stayed at a constant level. 
Usually, colour indices which show little changes with brightness indicate a stable circum-stellar disk \citep{2015A&A...574A..33R}. 
The saturation of V-I during the bright, long outburst may  suggest that the total emission from the star (bluer)  and the disk (redder) is completely dominated by the disk emission, maybe because the star is fully obscured by the disk.
A SALT-HRS spectrum as we have obtained it for the other two systems could could help to constrain the inclination and size of the disk to verify such a scenario. 

\subsection{\xmmb}
\label{subsec:discub}

\sxmmb was in the FoV of two pointed \xmm observations.
The factor of 5--10 higher luminosity compared to \sxmma allowed us to do a detailed spectral and temporal analysis.
The X-ray spectrum is well represented by an absorbed power law with photon index 0.6, which is on the lower end of the distribution \citep{2004A&A...414..667H} and indicates a harder spectrum. The X-ray luminosity between the two \xmm and other observations back to 1993 with \rosat changed by a factor of about two, while on shorter time scales of hours, similar variations are seen.

We discovered X-ray pulsations in the flux of \sxmmb with a period of $\sim$317.7\,s. The LS periodogram shows peaks at the fundamental frequency and three harmonics, caused by a highly complex pulse profile (Fig.\,\ref{fig:ppb}) with a deep narrow (0.1 in phase) dip.
Pulse profiles of Be/X-ray binary pulsars are known to be complex in soft ($<$10\,keV) X-ray energy ranges with
multiple emission peaks and sometimes narrow absorption dips, which go deeper than the broader minima between emission peaks. Phase-resolved spectroscopy revealed that matter in the accretion streams partially obscures the emitted radiation causing the dips \citep[e.g.][]{2012MNRAS.420.2307M,2013ASInC...8..103N}.
Examples in the Magellanic Clouds are 
Swift\,J053041.9$-$665426 = LXP\,28.8, which shows a narrow dip, very similar to \sxmmb, with a width of $\sim$0.1 in phase \citep{2013A&A...558A..74V}, but less well resolved due to a lower statistical quality of the pulse profile. 
This is also the case for XMMU\,J005929.0$-$723703 = SXP\,202b with possibly two dips \citep[widths of 0.1 and 0.13 in phase,][]{2008A&A...489..327H}.  

The dip seen in the pulse profiles of \sxmmb shows a peculiar phase dependence with energy. It is broader in the soft (0.2--2.0\,keV) band as compared to the hard (2.0--12\,keV) band, which results in an HR increase at the beginning and end of the dip. However, when the dip is deepest in the hard band, the HR shows a sudden drop. This resembles a partial eclipse of the X-ray source, which is surrounded by dense material that leads to increased absorption before and after the eclipse. An eclipse-like event could originate from a special viewing geometry when our line of sight passes parallel through the accretion column onto one of the magnetic poles of the neutron star. 
To further explore this scenario, data of higher statistical quality are required to allow a detailed pulse phase spectroscopy with sufficiently fine phase bins to resolve the dip and test more complex absorption models.

As optical counterpart we identify 2MASS\,\massb, a V=14.2\,mag star. Brightness and colours are typical for a Be star in the LMC. 
Optical spectroscopy with SALT HRS confirms the Be nature and revealed complex \Halpha and \Hbeta emission line profiles. From modelling of \Halpha line profiles of Be stars \citep{2010ApJS..187..228S} such complex structures are not expected. An axisymmetric disk is expected to produce a double-peaked, symmetric line profile, while asymmetric profiles are thought to arise  from  one-armed  density waves in the circum-stellar disk. These models were developed for isolated Be stars, the presence of a neutron star in a BeXRB likely leads to additional distortions of the disk and may be the reason for the higher complexity of the \Halpha emission observed from \sxmmb.

A simple modelling of the \Halpha profile with multiple Gaussian lines suggests four emission lines, one broad and three narrower (Table\,\ref{tab:hrs}), which indicate velocity differences of the emitting gas of up to $\Delta v$ = 125\,\kms along the line of sight.
The two lines, which are responsible for the double-peak appearance, are located relatively symmetric around the centre of the broad line, with $\Delta v \sim 66$\,\kms.
Assuming this is caused by Keplerian rotation, we can constrain the distance of the emitting gas to the central star. 
With a Keplerian velocity $v_{\rm K}$ = $\sqrt{GM/r}$, with $r$ the distance of the emitting gas to the central star with mass $M$ and $G$ the gravitational constant, we obtain 
\begin{equation}
r = \frac{G\,M\,sin^2i}{(0.5\,\Delta v)^2}, 
\label{eqt:rdisk}
\end{equation}
with $i$ the disk inclination \citep{2017MNRAS.464..572M}.
Adopting $M = 17\,M_\sun$ for a star with spectral type B0, typical for BeXRBs, results in $r = 2.1\,sin^2i \times 10^{12}$\,m.
This value, together with the relatively small equivalent width of -8\,\AA, which indicates a small disk, requires a small inclination angle  \citep[][]{2017MNRAS.464..572M,2006ApJ...651L..53G}.
The stronger blue-shifted line may be the result of a one-armed oscillation, with an enhanced density region on the blue end of the disk.
The additional line at \hbox{-90}\,\kms suggests that this region also extends to the inner disk ($r \propto 1/(\Delta v)^2$).
The \Hbeta line, which shows a very similar profile, is consistent with this picture. The derived $\Delta v$ of 100\,\kms results in a factor of 2.3 smaller distance from the star.

The OGLE I and V light curves show a very strong increase in brightness over about 850\,days, which is superimposed by regular outbursts (which repeat every $\sim$274\,days) with an amplitude of $\sim$0.15\,mag (in I). After the brightness reached a maximum of 13.1\,mag a sudden drop by $\sim$1\,mag occurred within 73--165\,days. Because the minimum was not sampled (gap of 92\,days), the drop by 1\,mag needs to be regarded as a lower limit.
After the drop the brightness recovered to its pre-outburst level.
A similar recovery was also recorded at the beginning of the OGLE monitoring, about 4395\,days (12\,years) before the one in Aug. 2013. Unfortunately, whether the first dip was also preceded by a similar outburst will remain unknown.

The behaviour of \sxmmb in the optical is very similar to an outburst from 3XMM\,J051259.8$-$682640 \citep{2017A&A...598A..69H}, however even more extreme in amplitude. 3XMM\,J051259.8$-$682640 showed three deep dips, which occurred 1350\,days (3.7\,years) apart, but only one of the three was preceded by an outburst. 
\citet{2017A&A...598A..69H} proposed that such regular dips might indicate the orbital period of the binary system and the dips originate from the interaction of the neutron
star with the circum-stellar disk of the Be star when the NS approaches
the Be star in an eccentric orbit. This could lead to a (partial) disruption of the disk, which is replenished until the NS returns. The more than three times longer dipping (orbital?) period in \sxmmb might allow the disk to obtain a larger size than in 3XMM\,J051259.8$-$682640, resulting in a stronger outburst and deeper dip.

However, the OGLE light curve of \sxmmb also exhibits a shorter periodicity around 274 days, which originates from three strong outbursts that occurred during the rise to the brightness maximum.
While this period is typical for binary periods of BeXRBs, it is not stable throughout the full light curve. Other smaller outbursts are not in phase with the stronger ones. 

Most of the time \sxmmb exhibited a linear correlation of V-I colour with brightness. In particular at the faintest level during the deep dip the system was bluest and reddened with brightness. 
Reddening with brightness is also observed from a number of BeXRBs in the Magellanic Clouds \citep[][]{2021MNRAS.503.6187T,2017A&A...598A..69H,2014A&A...567A.129V,2012MNRAS.424..282C}.
Such a behaviour is expected during disk build-up when more emission from the cooler disk is contributing. During the final rise to maximum brightness of \sxmmb the colour index reached an apparent limit, suggesting that the circum-stellar disk completely dominated the emission. 

\subsection{\rxjc}
\label{subsec:discuc}

The X-ray spectrum of the system may be fitted by an absorbed power-law component. However the presence of residuals in the high energies dictates the use of models with more free parameters. These residuals are often seen in BeXRBs in the MCs \citep[e.g.]{2014MNRAS.444.3571S,2020MNRAS.494.5350V}. Adding a high energy cut-off flattens the residuals in a more physical way than adding an additional thermal continuum component. Specifically we find that adding a black-body component with temperature of $\sim$1.95 keV and size of $\sim$460 m provides the best statistical fit. Although, such emission feature has been seen in persistent long period Galactic BeXRBs, the component is typically less hot (i.e.  $\lesssim$1.4 keV) and contributes a smaller percentage of the flux (i.e. $\lesssim$40\%) in the 0.3-10.0 keV band \citep{2013MmSAI..84..626L}.

As optical counterpart we identify a V=14.9\,mag star. The brightness, colours and the \Halpha emission seen in our SALT HRS spectrum, confirm the nature of \srxjc as a BeXRB. The \Halpha line profile is also structured with two broad and clearly resolved lines. Most likely, these originate from opposite sides of the circum-stellar disk and are shifted with $\Delta v$ of about 178\,\kms. 
Following Eqt.\,\ref{eqt:rdisk} with the same assumptions as for \sxmmb, we derive $r = 2.9\,sin^2i \times 10^{11}$\,m. \citet{2017MNRAS.464..572M} find a factor of about 10 smaller disk sizes from Galactic BeXRBs during epochs when their \Halpha line EQW was measured around -4\,\AA, similar to the value of \srxjc. This would indicate $i \sim 20\degr$.

The optical light-curve of \srxjc shows characteristic flares, however no clear periodic behaviour is evident. Long-term changes in optical magnitude are an ubiquitous characteristic of BeXRBs and are generally attributed to the Be disk build-up and disappearance. 
Optical flares might also mark the interaction of the NS and the Be disk during periastron passage. In fact given that the orientation of the disk can change due to precession, the resulting flares can also be quasi-periodic in nature as seen in the characteristic case of RX\,J0529.8$-$6556 \citep{2021MNRAS.503.6187T}, but also predicted by simulations \citep{2021ApJ...922L..37M}.

The colour magnitude diagram of \srxjc (see Fig. \ref{fig:oglecolc}) shows a characteristic hysteresis loop.
This loop-like track is evident in a number of BeXRBs in the MCs \citep[e.g.][]{2013A&A...554A...1M}, and in isolated Be stars as well \citep{2006A&A...456.1027D}.
This evolutionary track might be explained by a disk that grows during the rise, but becomes hollow during the decay forming thus a ring of somehow equal brightness but redder colour.

\section{Conclusions}

\begin{figure}
  \centering
  \resizebox{0.95\hsize}{!}{\includegraphics[clip=]{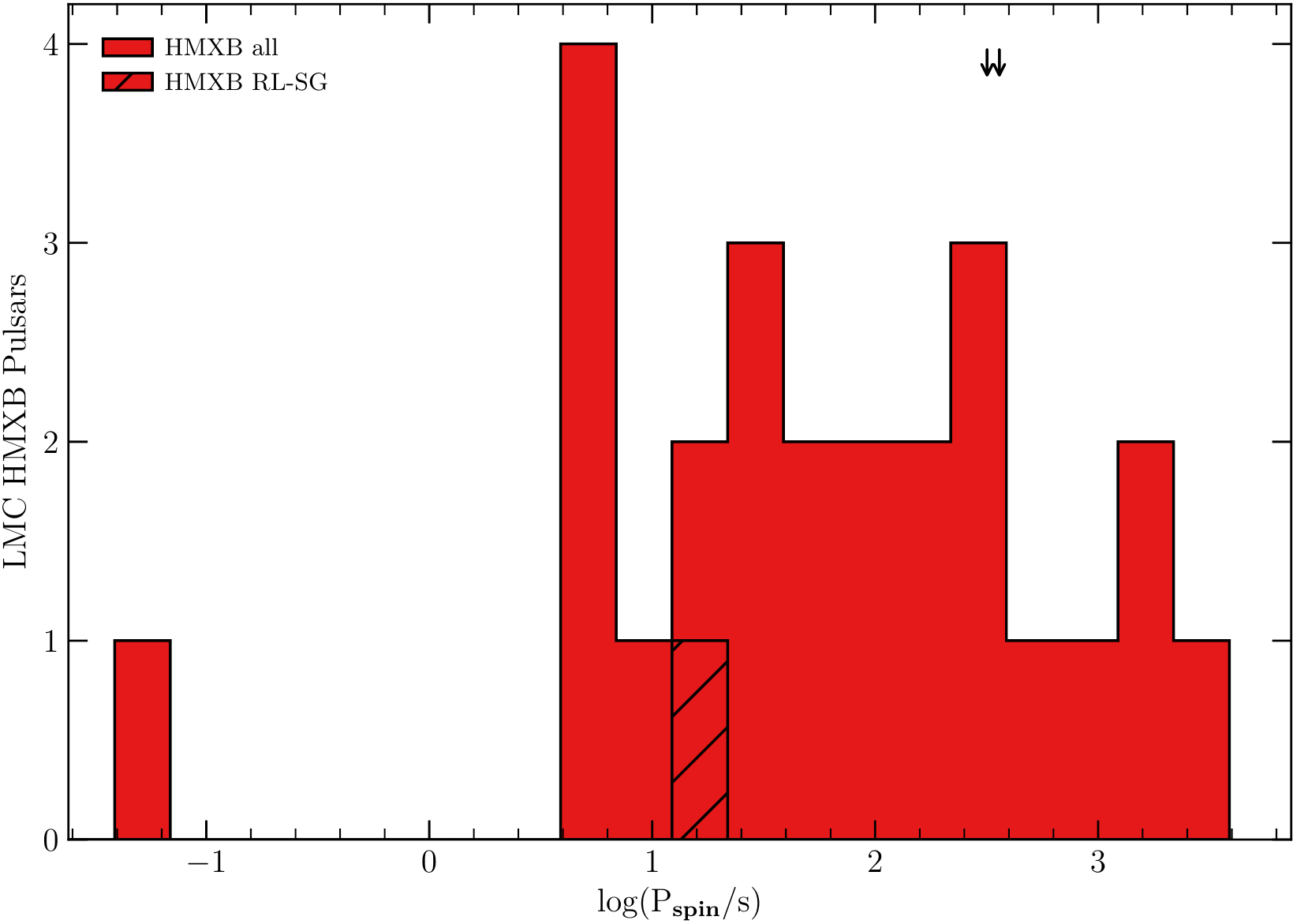}}
  \caption{Spin period distribution for 25 HMXB pulsars in the LMC. Apart from LMC\,X-4, which is a supergiant system powered by Roche-Lobe overflow, all other pulsars orbit a Be star. The two arrows mark the periods of the two newly discovered systems.}
  \label{fig:periods}
\end{figure}

Our discovery of two new BeXRB pulsars, with periods of 317.7\,s from \sxmmb and 360.7\,s from \srxjc, increases the number of known HMXB pulsars in our neighbour galaxy to 25. Their pulse period distribution is shown in Fig.\,\ref{fig:periods}.
\citet{2011Natur.479..372K} first reported a bimodal distribution of the spin periods of BeXRB binary pulsars in the LMC, the SMC and the Milky Way with two maxima at around 10\,s and several 100\,s, which are most clearly seen for the SMC systems \citep{2016A&A...586A..81H}. Although the number of LMC systems has increased considerably over the last years, there is no clear bimodal distribution visible yet from the LMC. 
Moreover, the LMC hosts a higher number of supergiant HMXBs and maybe a higher relative abundance of pulsars with NS spin periods longer than $\sim$1000\,s \citep[e.g.][]{2015A&A...579A.131C,2018MNRAS.475..220V,2021A&A...647A...8M}, indicating the presence of a younger population.
Continuation of targeted and serendipitous X-ray observations of the LMC will likely allow to uncover more of its HMXB pulsar population and allow more definite conclusions.

\bibliographystyle{aa} 
\bibliography{general} 

\begin{acknowledgements}
This work used observations obtained with \xmm, an ESA science mission with instruments and contributions directly funded by ESA Member States and NASA. The \xmm project is supported by the DLR and the Max Planck Society. 
This research has made use of the VizieR catalogue access tool, CDS,
Strasbourg, France. The original description of the VizieR service was published in A\&AS 143, 23.
This work has made use of data from the European Space Agency (ESA) mission
{\it Gaia} (\url{https://www.cosmos.esa.int/gaia}), processed by the {\it Gaia}
Data Processing and Analysis Consortium (DPAC,
\url{https://www.cosmos.esa.int/web/gaia/dpac/consortium}). Funding for the DPAC
has been provided by national institutions, in particular the institutions
participating in the {\it Gaia} Multilateral Agreement.
\end{acknowledgements}

\end{document}